\documentclass[runningheads]{llncs}
\usepackage{pgfplots}
\pgfplotsset{compat=1.18} 

\usepackage{enumitem}

\usepackage{adjustbox}
\usepackage{float}
\usepackage{multirow}
\usepackage[section]{placeins}
\usepackage{pgf-pie}  
\usepackage{array}
\usepackage{xurl}
\usetikzlibrary{mindmap}

\definecolor{ibmyellow}{HTML}{ffb000}
\definecolor{ibmorange}{HTML}{fe6100}
\definecolor{ibmmagenta}{HTML}{dc267f}
\definecolor{ibmindigo}{HTML}{785ef0}
\definecolor{ibmblue}{HTML}{648fff}

\usepackage{hyperref}
\hypersetup{
    colorlinks=true,
    linkcolor=ibmmagenta,
    filecolor=ibmorange,      
    urlcolor=ibmblue,
    citecolor=ibmmagenta,
    }

\usepackage[T1]{fontenc}
\usepackage{graphicx}
\usepackage{color}

\begin{document}
\title{Decentralization: A Qualitative Survey of Node Operators}
%
%
\author{
Alex Lynham\inst{1}\orcidID{0009-0005-0488-7651} 
\and
Geoff Goodell\inst{1}
}
\authorrunning{A. Lynham and G. Goodell}
%
\institute{$^1$University College London, 66-72 Gower St, London WC1E 6BT, England}
\maketitle              
\setcounter{footnote}{1}

\begin{abstract}
Decentralization is understood both by professionals in the blockchain industry and general users as a core design goal of permissionless ledgers. However, its meaning is far from universally agreed, and often it is easier to get opinions on what it is not, rather than what it is. In this paper, we solicit definitions of `decentralization' and `decentralization theatre' from blockchain node operators. Key to a definition is asking about effective decentralization strategies, as well as those that are ineffective. Malicious, deceptive or incompetent strategies are commonly referred to by the term `decentralization theatre.' Finally, we ask what is being decentralized. Via thematic analysis of interview transcripts, we find that most operators conceive of decentralization as existing broadly on a technical and a governance axis. This informs a two-axis model: \textit{network topology} and \textit{governance topology}, or the structure of decision-making power. Our key finding is that `decentralization' alone does not affect ledger immutability or systemic robustness.

\end{abstract}

\newcommand{\blockquote}{\medskip \noindent \leftskip 16pt}

\newcommand{\blockquotesmallskip}{\smallskip \noindent \leftskip 16pt}

\newcommand{\blockquoteendsmallskip}{\smallskip \leftskip 0pt}

\newcommand{\blockquoteend}{\medskip \leftskip 0pt}

\newcommand{\blockquoteendnoskip}{\leftskip 0pt}

\section{Introduction}
In this paper, we define `decentralization,' and analyse whether or not it has any impact on the rewrite resistance of permissionless ledgers. We advance a model of decentralization based on thematic analysis of fieldwork, finding that the \textit{governance topology} of a network takes precedence over its \textit{network topology}. Metrics in a ledger's network topology can affect rewrite resistance, but increasing decentralization of its governance topology has no effect. This means that, by our definition, increasing decentralization does not affect ledger resilience. This paper's research contribution is:

\blockquotesmallskip
\textit{advancing a model of decentralization that matches participants' experience of governance interacting with the technical topology of a network.}

\blockquoteendsmallskip

Decentralization, while difficult to define, is a core tenet of blockchain technology, and is defined via protocol-enforced incentives and technical means as well as by non-technical measures. We examine ways of understanding decentralization from existing literature in Section 2.3, and document the views of node operators on what it means in the real world in Section 3. Throughout, we address the views of node operators on decentralization, the responsibilities of validators, and their role within the organisational dynamics of a network. 

First, we advance a two-axis analytical model of decentralization in Section 4.1. We then theorize that, contrary to a general assumption by node operators that increased decentralization results in increased rewrite resistance, different network and governance topologies result in different manifestations of ledger rewrites (Section 4.3 and Section 4.6), but do not change their likelihood. For example, consortium ledgers might feature the risk that an authority uses its position to force change, and permissionless ledgers might be at risk of takeover by foundations, developer groups, or stakeholder blocs. Crucially, our breakdown of the two axes of decentralization, \textit{network topology} and \textit{governance topology}, that can be analysed both ex-ante and ex-post relies on a simple heuristic, asking the question `decentralization of what?'

To make the implicit explicit, we offer a model for roles in the typical structure of a Proof-of-Stake blockchain (Section 4.4). We use our interviews to inform the identification of personas, and support analysis of the reported dynamics between them. Though we focus on the Cosmos Ecosystem,\footnote{The Cosmos Ecosystem is an interconnected network of blockchains built using the same Software Development Kit, or SDK. By default, they are linked by the IBC protocol, and are Delegated Proof-of-Stake, built around a DAO (Decentralized Autonomous Organisation) for voting on protocol upgrades and governance issues.~\cite{whatiscosmos}} due to our recruitment methodology, much of our analysis can be applied to any network that uses Proof-of-Stake, or specifically, Delegated Proof-of-Stake,\footnote{See Bashir~\cite{Bashir2022} for a definition, and Section 4.4 for analysis.} as its security mechanism, and has governance and network topologies to those described by our respondents.

\section{Methodology and Context}
\begin{table}[t!]
    \begin{adjustbox}{width={0.98\textwidth},totalheight={\textheight},keepaspectratio}
\begin{minipage}[t]{.65\textwidth}
{\renewcommand{\arraystretch}{1.5}
\quad{\setlength{\tabcolsep}{0.4em}
\centering
    \caption{Survey and semi-structured interview recruitment \strut}
    \label{table:recruitment}
    \begin{tabular}{| p{0.25\linewidth} | p{0.25\linewidth} | p{0.25\linewidth} | p{0.25\linewidth} |}
         \hline
          \raggedright\textbf{Artifact}\arraybackslash& \raggedright\textbf{Number of recruitment forums}\arraybackslash& \raggedright\textbf{Estimated Reach}\arraybackslash& \raggedright\textbf{Participants}\arraybackslash \\
         \hline
         Survey&   5&500& 35\\
         \hline
         Semi-structured interviews&   6&550& 7\\
         \hline
    \end{tabular}
}}
\end{minipage}%
\qquad\qquad
\begin{minipage}[t]{.65\textwidth}
{\renewcommand{\arraystretch}{1.5}
\quad{\setlength{\tabcolsep}{0.4em}
\centering
    \caption{Number of chains validated by node operators' organisations \strut}
    \label{table:chains_validated}
    \begin{tabular}{| p{1\linewidth} | }
         \hline
          \textbf{Number of Chains validated} \\
         \hline
         About 15 \\
         \hline
         More than 20 \\
         \hline
         83 \\
         \hline
         About 50 \\
         \hline
         About 100 \\
         \hline
         About 20 \\
         \hline
    \end{tabular}
}}
\end{minipage}
    \end{adjustbox}
\end{table}
We undertook an anonymous survey and semi-structured interviews of node operators from organisations that run validator nodes. Blockchains, whether Proof-of-Work, or (delegated) Proof-of-Stake, require nodes to operate, and consequently, we have focussed on the technical practitioners that operate these nodes for our fieldwork, rather than a more general survey of stakeholders. We recruited from private validator forums and channels, as well as Foundation or Core Team invite-only managed channels for active validators, with the teams' permission. The exact channels are not revealed here so as to avoid the possibility of identifying an interview participant, or their organisation, based on their publicly observable network participation.

Due to the potentially serious consequences in terms of reputational damage, impact to businesses or even threats to individuals, we could only proceed under a protocol designed to ensure strict anonymity of participants. For the survey, this meant an anonymous approach, and for the semi-structured interview, this meant immediately transcribing the conversation, before deleting the source audio and assigning it a UUID.

For both survey and interviews, the goal was to target technical members of staff; however, for reasons of anonymity, we did not ask for interviewees' roles in case that became a data point usable for de-anonymising a participant. For similar reasons, we did not collect demographic data about respondents.

It is hard to establish the unique operators in a given set of channels, since many operate on several chains, and multiple staff from a single company may be in a channel, but we estimate the reach of our recruitment to have been in the hundreds. Most channels do not expose public membership numbers or lists, so this estimate is based on the size of mainnet validator sets for the channels maintained by networks. For context, using node data collected on the 3rd of May 2024, across 62 Cosmos chains, there were 2,275 unique validators.\footnote{A time-stamped image of the chain explorer Mintscan can be found in \url{https://github.com/envoylabs/cosmos\_data}; the raw data is not exposed via API.} The data are summarized in Table \ref{table:recruitment}.

\subsection{Survey}
An anonymous survey was posted to private validator communication channels, private threads and private validator forums for networks. Typically, these threads are kept up-to-date by Core Teams and Foundations,\footnote{For a definition of these agents, see Table \ref{table:agents} in Section 4.} and they are invite-only. The reason is that these channels are used for coordination in the event of a security patch, or cyber attack. Participants in these operational channels are generally technical staff, but could also be non-technical staff from node operator organisations. 35 surveys were filled out, including 14 partial responses.

\subsection{Semi-structured Interviews}
We recruited respondents from the same channels, plus an additional channel outside of Cosmos, to take part in a semi-structured interview. Each interview began with a script, then a warm-up question on number of chains validated (summarized in Table \ref{table:chains_validated}),\footnote{Not every respondent gave an on-record answer for this.} followed by the discussion areas in Appendix C. These interviews used discussion topics, with the interviewer prompting for clarifications where needed. After transcription, the source audio was deleted and a UUID assigned to the transcription. Seven operators agreed to take part, following initial concern that no volunteers would come forward.

\subsection{Decentralization}
The meaning of ‘decentralization’ is far from universally agreed. The ISO definition of a decentralized system (ISO 22739:2024 3.20)~\cite{isoblockchain} is a start, a ``distributed system (3.24) wherein control is distributed among the persons or organizations participating in the operation of the system.'' This raises the question of how that control is distributed, which has become a key question for regulators.\footnote{The EU's MiCA cryptoasset regulation explicitly uses decentralization as a test, ``Where crypto-asset services are provided in a fully decentralised manner without any intermediary, they should not fall within the scope of this Regulation.''~\cite{mica}} The ISO definition simply notes, ``the distribution of control among persons or organizations participating in the system is determined by the system’s design.''

Gochhayat et al.,~\cite{gochhayat} conceive of a blockchain system as consisting of a governance layer, a network layer, and a storage layer, and analyse the degree of decentrality at each. Like node operators, as we will see, they assume that better decentralization results in greater resilience:

\blockquote
``Decentralized systems are a subset of distributed systems where multiple authorities control different components and no authority is fully trusted by all. Decentrality is a property related to the control over the system. Better decentralization means higher resistance against censorship and tampering.''

\blockquoteend
 
Of the three layers they enumerate, the governance layer is the most important for our analysis, relating more strictly to voting power in the protocol, in their case Proof-of-Work, by ``nodes participating in the block creation or mining process'' and their hash rate. This is similar to other definitions such as Lin et al.,~\cite{linetal9438771} which also focus on distribution of mining power. Gochhayat et al.~\cite{gochhayat} also advance definitions of network and storage layers that are useful for our analysis. The network layer involves control of network topology, bandwidth and block dissemination, factors more important in probabilistic finality, and less important in a Delegated Proof-of-Stake network with absolute finality. The storage layer meanwhile relates to node location, datacentre or cloud provider, which maps closely on to definitions of decentralization given in our fieldwork by node operators, who mentioned hardware and geographic dispersion of nodes.

A better definition might be a multi-faceted one, such as that advanced by JP Vergne. Vergne attacks decentralization from a different angle, identifying theatrical modes for the discussion. ``I propose to view the Web3 community as a decentralization theater,'' he suggests, ``namely, a fictitious space whose performers are enjoined to enact decentralization to preserve the coherence of the play.''~\cite{vergne} He gives a terse definition of decentralization, which we will return to when discussing Figure \ref{fig:naive_decentralization} in Section 4.3; ``decentralization concerns authority dispersion within an organization, a system, a digital platform, or a network (typically, blockchains are these four things all at once).''~\cite{vergne} Although his modes of ``decentralization theatre'' are satirical, they're memorable enough names that we will adopt them unironically in our analysis:

\begin{enumerate}[noitemsep]

\item \textbf
{Bucket Decentralization} This is a common discourse in the blockchain space, where decentralization is framed in economic terms, ``equating economic inequality with token ownership concentration and, by extension, with centralization.''~\cite{vergne} Its logical failings are discussed at length by Vergne, but for our purposes we understand it to mean \textit{economic decentralization, particularly that concerned with distribution of voting power} (i.e. staked tokens).
\item \textbf{Godot Decentralization} This is another common framing of decentralization, broadly speaking referring to \textit{systemic resilience via technical means}, ``[f]or some, a system is decentralized when it has so many points of failure that it simply cannot fail amid adverse circumstances, for any attacker would have to simultaneously attack on too many fronts to be successful.''~\cite{vergne}

\end{enumerate}

To begin, we will adopt a definition of decentralization from an interview participant,\footnote{Extracts from our transcripts are in the Appendix. Quotes from our transcripts not in the Appendix bear the anonymous common citation `interview participant.'} ``there is no one entity that can control the state of information being held by the blockchain,'' and add our hypothesis: ``Decentralization, expressed in terms of control of ledger state, is a holistic expression of the governance layer, that is, on-chain voting power, off-chain coercion, and social influence.'' 

In the spirit of Vergne’s work, actors in the system constitute it and can subvert it - all the world’s a protocol, and all the men and women merely players.

\section{Results}
What follows are summary insights from our semi-structured interviews and survey free-text boxes. The transcripts of discussions on decentralization from the semi-structured interviews, and survey responses can be found in the Appendix. Tables \ref{table:results_coding1} and \ref{table:results_coding2} show a process of thematic analysis, inductively coding into themes, and then aggregating into two topologies that are discussed in greater depth in Section 4. Example variables from node operators are enumerated in tables \ref{table:network_topology} and \ref{table:governance_topology} along with common metrics from prior literature, and deductive metrics validated by the interviews.

Principally, decentralization was defined by interviewees in terms of an absence of central control, especially over ledger state. Discussions of control are linked to token distribution, as a proxy for control of decision-making. A key insight is that many ledgers have a de facto central authority, exercising power via subsidy programmes like Foundation delegations for validators and developers. Like in the Gochhayat definition, there is a pattern to responses that conflates systemic robustness or censorship resistance with `decentralization.'

Operators often described decentralization as a multi-layered system of coercion resistance akin to Vergne's `Godot Decentralization.' In quantifying it, geographic distribution was often mentioned, as was voting power across nodes. Operators typically mentioned hardware and geographic distribution of nodes first. Another repeated theme was that of performativity, which was foregrounded when explicitly discussing `decentralization theatre.'

{\renewcommand{\arraystretch}{1.5}
\quad{\setlength{\tabcolsep}{1em}
\begin{table}[t!]
    \begin{adjustbox}{width={\textwidth},totalheight={\textheight},keepaspectratio}
    \begin{minipage}[t]{.62\textwidth}
    \caption{Thematic analysis (a): 1 of 2\\ \textbf{Aggregate Theme}: Governance and ownership of the network}
    \label{table:results_coding1}
    \begin{tabular}[t]{|p{0.8\textwidth}|>{\raggedright\arraybackslash}p{0.2\textwidth}|}
         \hline
         \raggedright \textbf{Property of `decentralization' or `decentralization theatre'}\arraybackslash & \textbf{Theme} \\
         \hline
         ``The idealistic definition is\ldots no one entity that can control the state of information being held by the blockchain, so that there's not a risk of an authoritarian coming in and rewriting history'' (Appendix D, `Decentralization', Response 1)&  No central control of ledger state\\
         ``there's no single party controlling the chain.'' (Appendix D, `Decentralization', Response 3)&  \\
         ``it's not run by a select few individuals, or a specific company.'' (Appendix D, `Decentralization', Response 5)
&  \\
         ``Everything not being controlled by a single entity'' (Survey response)
&  \\
         ``A substantial effort is made to avoid the system being dependent on a few centralised entities.'' (Survey response)
&  \\
         \hline
         ``the token controlling interest is held by a select few, sort of like a dev team, and they have a Foundation that is delegating votes, then in effect they have the controlling interest.'' (Appendix D, `Decentralization', Response 1)&  Founda-tion token ownership and delegations\\
         ``[Foundations] in most cases have a very large amount of tokens, and they can use that to either bring people into the validator set, or push them out,'' (Appendix D, `Decentralization', Response 5) &  \\
         ``I would say that most blockchains today are quite decentralized, even though some people will say, you know, Foundations have twenty percent of stake. Well, that's fine'' (Appendix D, `Decentralization', Response 2)&  \\
         ``there are a number of chains that do this, and all of them do it to some extent\ldots controlling decisions via stake, and so on.'' (Appendix E, `Decentralization Theatre', Response 2)&  \\
         \hline
         ``even though nodes can be geographically dispersed, I've found that in most cases the majority of the stake is still owned by a small party that could, in fact, enact any change they want.'' (Appendix D, `Decentralization', Response 6)& Voting power concentration \\
         ``decentralization of tokens and voting power.'' (Appendix D, `Decentralization', Response 3)&  \\
         ``Having a decentralized system (decentralized validator set) that is effectively controlled by a single entity,'' (Survey response)&  \\
         ``Appearing to be decentralised when in reality the system is either run by or controlled by a select few,'' (Survey response)
&  \\
         \hline
         ``Nakamoto coefficient is probably the most common metric, but that's pretty well decentralization theater.'' (Survey response)&  Perform-ative decentralization\\
         ``Everybody wants to be decentralized, until it affects their bottom line'' (Appendix E, `Decentralization Theatre', Response 5) &\\
         ``decentralization as a talking point, but not an actual goal.'' (Survey response)&  \\
         \hline
    \end{tabular}
    \end{minipage}%
    \qquad\qquad\qquad\qquad
    \begin{minipage}[t]{.62\textwidth}
    \caption{Thematic analysis (b): 2 of 2\\ \textbf{Aggregate Theme}: Physical and organisational structure of the network}
    \label{table:results_coding2}
    \begin{tabular}[t]{|p{0.8\textwidth}|>{\raggedright\arraybackslash}p{0.2\textwidth}|}
         \hline
         \raggedright \textbf{Property of `decentralization' or `decentralization theatre'}\arraybackslash & \textbf{Theme} \\
         \hline
         ``each validator is run by a different person or company, and that validator is on slightly different hardware. Perhaps it's also a different operating system, and on different fibre-optic carriers.'' (Appendix D, `Decentralization', Response 4)&  Collusion resistance and operator independence\\
         ``There's multiple different related parts that are involved\ldots If any of them are compromised, then the whole system doesn't work.'' (Appendix D, `Decentralization', Response 3)&  \\
         ``every action is spread out to multiple participants'' (Survey response)&  \\
         ``Enough parties involved that the ability to collude to change state, steal, coerce, manipulate is impractical/impossible.'' (Survey response)&  \\
         ``a diverse set of independent service operators who operate following a set of published rules.'' (Survey response)&  \\
         \hline
         ``every validator on a particular blockchain is somehow evenly distributed across all the different countries in the world, so you have one validator in every single country'' (Appendix D, `Decentralization', Response 4)&  Geogra-phical distribution of nodes\\
         ``geographic decentralization of the validators running a chain" (Appendix D, `Decentralization', Response 3)&  \\
         ``It could be hosting providers, it could be machines all running in the same country, or the same timezone, for example, and it could also be where the citizenship or the actual company is based.'' (Appendix D, `Decentralization', Response 7) & \\
         ``Geographically distirbuited [sic] and not tied to a single organization (or a small accuntable [sic] number of them)'' (Survey response)&  \\
         ``Enough geographic distribution that network/geographic/acts-of-god situations don't impact liveness or state'' (Survey response)&  \\
         ``Having as many nodes geographically and provider diverse as possible on your own bare metal hardware located on private property that you own.'' (Survey response)&  \\
         ``a distributed network not only through distributed VP but also location and hosting providers.'' (Survey response)&  \\
         ``There are a handful of entities controlling most chains, and they're almost all mostly in the same Contabo or Hetzner data centers.'' (Survey response)&  \\
         \hline
    \end{tabular}
    \end{minipage}
    \end{adjustbox}
\end{table}
}}

\section{Analysis}

\subsection{The Two-Axis Decentralization Model}

\begin{figure*}[ht!]
\centering
\begin{adjustbox}{width={\textwidth},totalheight={2.5in},keepaspectratio}
\begin{tikzpicture}
  \path[mindmap,concept color=ibmmagenta,text=white,
  every annotation/.style={fill=ibmyellow}
  text width=2cm, align=flush center
  level 1/.style={level distance=10cm},
  level 2/.style={level distance=3cm}
 ]
    node[concept] {Decentralization}
    [clockwise from=-60]
    child[concept color=ibmorange] {
      node[concept] (gov) {Governance Topology}
      [clockwise from=70]
      child { node[concept] {Distribution of staked tokens} }
      child { node[concept] {Number of proposers of governance votes} }
      child { node[concept] {Participation in on-chain governance votes} }
      child { node[concept] {Consensus mechanism (e.g. PoW, PoS, DPoS) } }
    }
    child[concept color=ibmblue] {
        node[concept] (net) {Network Topology} 
        [clockwise from=-70]
      child { node[concept] {Validator node locations} }
      child { node[concept] {Validator node jurisdictions} }
      child { node[concept] {Number of client implementations} }
      child { node[concept] {Number of validators} }
    };
\end{tikzpicture}
\end{adjustbox}
\caption{Two-axis Decentralization Model with example metrics}
\label{fig:two_axis}
\end{figure*}
We advance a two-axis model for defining decentralization, shown in Figure \ref{fig:two_axis}. It is divided between \textit{Network topology} (the jurisdictional and physical structure) and \textit{Governance topology} (the structure of decision making power). We argue that decentralization is an emergent, runtime property of the protocol.

If there exists a concise definition of `decentralization,' it should be a measure of the dispersion of decision-making power, closely mirroring the ISO definition (see Section 2.3), applied to individual variables within the topologies we describe. This becomes problematic only when the dispersion of decision making power is equated with systemic robustness or immutability (as in Fig. \ref{fig:naive_decentralization}, whereas we argue in Table \ref{table:leviathan_table_1} that this can never be guaranteed; only the manifestation of a rewrite will change under different structures of power), hence the need for a holistic, qualitative approach.\footnote{For instance, when considering an individual variable such as token distribution, it matters not only what the distribution is, but whether or not those tokens are staked; thus the consensus mechanic matters, and if stake can be bonded, so does who they are staked \textit{to}. In other words, few of the variables are independent.}

Key to our model offering clarity on decentralization is asking another question, which we think of as \textit{Vergne's Razor}: `decentralization of what?' Asking this question gives us the variables we need to investigate. We can then use this heuristic to categorize many aspects of the design and operation of heterogenous networks before-the-fact. To arrive at our two-axis model, we build on the existing literature, noting that threshold-based checks on individual variables, such as the Nakamoto Coefficient~\cite{srinivasan_lee} or Edinburgh EDI MDT~\cite{edi_ovezik2024sokstratifiedapproachblockchain}, are compatible with our framing. Other example metrics, such as the consensus mechanic, are categories. 

It is key to the argument of this paper that any definition of decentralization that can yield a quantitative measure for decentralization is likely performative at best, and misleading at worst. This is especially true if it is designed to be used for regulation. Even in the case of metrics that are as clear-cut as the Nakamoto Coefficient, a holistic view is needed. Inequality measures\footnote{Such as the Gini Coefficient or HHI, often used for antitrust metrics.~\cite{hhi_antitrust}} applied to blockchains that yield statements such as ``Network X is 0.7 Decentralized'' are rightly looked on with suspicion, as they usually only interrogate one, or a small number of highly observable, and manipulable, variables. For the purposes of analysis, \textit{Vergne's Razor} is informative, but ultimately, these are just single variables within the topologies we describe.


Unlike other models, we see decentralization as an expression of metrics and outcomes rather than an end goal in itself. In Section 4.3 we discuss this in the context of a common assumption by literature, such as Gochhayat,~\cite{gochhayat} and some node operators (see Section 3) that the goal of decentralization is systemic resilience, expressed as ledger immutability. These are not the same. Our model is designed for analysis and critique; where it has ex-ante power is in designing systems with novel topologies to those in existence.

Governance topology neatly encapsulates power, both political and economic, as well as the social dynamics of the agents in the system, and it is possible to interrogate the decentralization of network topology on different networks, and then compare them. Unlike the harder concerns of node operators, variables like hardware, client, and datacentre, analysis of such topologies can cover the operators' organisations and their physical jurisdictions, which is likely to be as important, if not more important, than where they run their servers.

It is not easy to be exhaustive about every variable, but most commonly expressed measures in our research fall into one of the two topologies (see Appendix B). These variables include metrics that our fieldwork shows are considered ``decentralization theatre'' if they are offered as a silver bullet, like the raw number of validator nodes (Appendix E, `Decentralization Theatre', Response 3). Much of the data required to measure these variables is publicly exposed to the network.

\subsection{Compatibility with Existing Definitions of Decentralization}

Many common definitions of decentralization refer to at least two characteristics:

\begin{enumerate}[noitemsep]
    \item distribution of \textit{nodes} (geographically, by datacentre, by OS/hardware, etc.)
    \item distribution of \textit{tokens}
\end{enumerate}

The first of these can be linked to Vergne's `Godot Decentralization.' The second, token distribution, can be linked to `Bucket Decentralization.' As described by Vergne, both are theatrical in the sense of being performative. 

Comparing this to Gochhayat et al.'s three layer model,~\cite{gochhayat} the distribution of tokens broadly corresponds to the governance layer, while the distribution of nodes fits within the storage layer. The network layer sits less easily, but questions of bandwidth and dissemination are likely to fall into the same technical category as the distribution or location of nodes. 

The ISO definition in Section 2.3 is better in one crucial regard: that decentralization is considered independent of resilience to rewrite. This contrasts with the general assumption of operators in our fieldwork. The Nakamoto Coefficient, ``the minimum number of entities in a given subsystem required to get to 51\% of the total capacity\ldots [or] the operative threshold''~\cite{srinivasan_lee} that is, 33.4\% of voting power in Cosmos networks, is another inadequate definition. It was mentioned in our fieldwork in the context of defining `decentralization theatre', ``when a chain has a high Nakamoto Coefficient but in reality there is 70\% of the chains [sic] VP running on the same hosting provider as a single point of failure.''\footnote{See Appendix D, Response 3 for another discussion of this.}

The Edinburgh Decentralisation Index (EDI) represents one of the most exhaustive, multi-faceted empirical metric and reporting tools available.~\cite{edi} As in our fieldwork, and the Nakamoto Coefficient article,~\cite{srinivasan_lee} they identify variables such as client diversity, number of software developers committing to core repositories, token holding and voting power concentration. However their approach is quantitative, whereas ours is qualitative. Moreover, they privilege geographical decentralization as a meta layer,~\cite{edi_ovezik2024sokstratifiedapproachblockchain} whereas we find that the governance topology of a network takes precedence. Geographical dispersion of nodes is just one variable among many in the blockchain's network topology.

Nevertheless, our analysis has much in common. We also attack the issue from a cybersecurity perspective, and are concerned with the same core design properties, ``safety'' and ``liveness.'' In future work, we intend to explore the question of whether cryptoassets can achieve the other design goals identified, ``price stability'' and ``privacy.''\footnote{We argue that public visibility renders cryptoassets more akin to a simplified form of credit, lacking the price stability of cash, as the security mechanism of these networks, economic security, relies on the ``collapse'' of the asset, as identified by Nakamoto in the Bitcoin whitepaper~\cite{bitcoin_nakamoto} and detailed by Budish et al.~\cite{budish2024economiclimitspermissionlessconsensus}} Finally, we would hypothesise that our model is also stratified on public permissionless networks, for a network's governance topology can subvert its network topology. Indeed, it is notable that the bottom four layers in their stratified model (Hardware, Software, Network and Consensus) would fall within network topology, and two of the top three layers (Economics and Governance) within governance topology. The more fully the network topology is visible and known, the more fully it can be targeted. The same is true for the agents that constitute the network.

Simplified heuristics such as the Nakamoto Coefficient, or the Edinburgh Decentralization Index's Minimum Decentralization Test\footnote{``A blockchain system fails the Minimum Decentralization Test (MDT) if and only if there exists a layer for which there is a single legal person that controls a sufficient number of relevant parties so that it is able to violate a property of interest.''~\cite{edi_ovezik2024sokstratifiedapproachblockchain}} are not granular enough to be useful in assessing anything other than whether a blockchain is at immediate, trivial risk of takeover.\footnote{A development of this idea can be found in our companion paper, where we describe the goal of immutability in terms of prevention of opportunistic actions.~\cite{lynham2025definingdltimmutabilityqualitative}} The MDT is better than Nakamoto in one key respect; that it can capture more subtle, coercive situations. Its more abstract definition means that where Nakamoto requires a direct link between an agent and their control of the ``operative threshold'', the MDT addresses a more subtle interplay of incentives and risk for the agents involved.\footnote{To summarize an example of this, in a DPoS network where validators vote stakers' shares if they neglect to vote, a Foundation or large holder (`Whale') can become the biggest single delegator to a number of validators with large voting power. As their largest delegator, these validators may become de facto controlled by the Foundation or Whale, thus bringing with them the votes of any stakers that do not vote, and magnifying their control of voting power beyond their strict percentage holding. (Appendix D, Response 7)} For this reason, the MDT applied to individual metrics within our topologies is superior to Nakamoto for simple analyses. Elsewhere, more complex definitions such as Axelsen et al.'s ``sufficient decentralization'', defined as ``a verifiable state, where (1) the design of the DAO is collusion resistant and based on long-term equilibrium; (2) its governance processes have unrestricted and transparent access,''~\cite{Axelsen_2022} or the full EDI do not mesh with participants' experience or describe a stable equilibrium for blockchains that can be used as a base for policy discussion, or ex-ante design. Axelsen et al observe that some argue ``no large-scale social, economic or political institution can be fully decentralized and automated without human intervention,''~\cite{Axelsen_2022} which neatly summarizes our argument, provides rationale for the two-axis model, and motivates our companion paper on conditional immutability, trust and legitimacy in ledger governance.~\cite{lynham2025definingdltimmutabilityqualitative}

\subsection{Decentralization and Immutability}
\begin{figure}[t!]
\centering
\begin{tikzpicture} 
\begin{axis}[
    height=2.4in,
    width=2.4in,
    ytick={0, 1},
    yticklabels={0, 1},
    xtick={0, 1},
    xticklabels={Low, High},
    xlabel=Decentralization,
    ylabel=Probability of rewrite,
]
    \addplot [mark=none,  ibmmagenta,   ultra thick] coordinates { (1,0) (0,1)};
\end{axis} 
\end{tikzpicture}
\caption{Naive model of decision making power vs immutability}
\label{fig:naive_decentralization}
\end{figure}
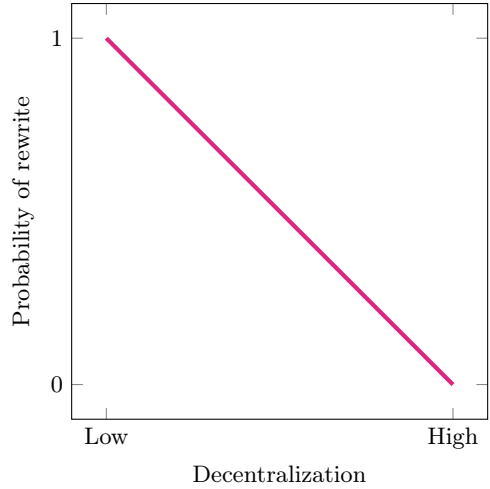

We might intuit, as both agents and authors like Gochhayat et al.~\cite{gochhayat} appear to, that greater `decentralization', or dispersion of decision-making over ledger state,\footnote{Dispersion of decision-making power recalls Vergne's definition of decentralization as dispersion of authority: ``authority [consists] of two components: (1) The ability to access information and (2) the ability to contribute to decision-making.''~\cite{vergne}} either adding to it, or governing it, would result in greater likelihood of future immutability (Figure \ref{fig:naive_decentralization}; see Gochhayat et al.'s link between decentralization and censorship resistance, and our initial definition of decentralization in Section 2.3). However, we hypothesise that as decision-making dispersion increases, on either the network or governance axis, security does not increase, but the form of a rewrite changes (Table \ref{table:leviathan_table_1}). This brings us full-circle to the question of whether all `decentralization' is performative, if its stated goal is to secure ledger state.\footnote{Contrast Figure \ref{fig:naive_decentralization} with self-reported validator participation in rewrites in Figure \ref{fig:q8}.}

Both in the case of regulation and the motivations of validators and network participants it is important to interrogate the objective served by increased decentralization. For both network participants and regulators, the answer appears to be immutability first, and then dispersion of decision making second. We examine the case of immutability more extensively in our companion paper on conditional immutability,~\cite{lynham2025definingdltimmutabilityqualitative} but for now it is enough to say that these networks are not immutable, regardless of their dispersion of decision making power (see Table \ref{table:leviathan_table_1}). We briefly summarise several rewrite case studies mentioned in our interviews in Table \ref{table:rewrites}. If the goal of greater decentralization is immutability, then we must consider not how to increase decentralization, but which variables affect ledger immutability when they are modified, and how they interact.

\begin{table}[t!]
    {\renewcommand{\arraystretch}{1.5}
\quad{\setlength{\tabcolsep}{1em}
    \centering
    \caption{Takeovers, coercion and collapse on permissioned and permissionless ledgers}\label{table:leviathan_table_1}
    \begin{adjustbox}{width={\textwidth},totalheight={\textheight},keepaspectratio}
    \begin{tabular}[t!]{|p{0.2\linewidth}|p{0.4\linewidth}|p{0.4\linewidth}|} \hline 
         \textbf{Dispersion of decision making power}&  \textbf{Permissioned ledger}& \textbf{Permissionless ledger}\\ \hline 
         \textbf{High}&  \textbf{Coherence collapses:} In a permissioned context the highest dispersion will still leave the authority a veto (e.g. 51\% to 49\% of voting power). Users cannot force change as authority can veto or revert; thus users can only leave& \textbf{Takeover:} Users force change\\ \hline 
         \textbf{Medium}&  \textbf{Action by Authority:} Authority and stakeholders force change& \textbf{Takeover:} Foundation, DAO, Validators, or large token holders force change\\ \hline 
         \textbf{Low}&  \textbf{Action by Authority:} Authority forces change& \textbf{Takeover:} Foundation or controlling DAO forces change\\ \hline
    \end{tabular}
    \end{adjustbox}
    }}
\end{table}
\begin{table}[t!]
{\renewcommand{\arraystretch}{1.5}
\quad{\setlength{\tabcolsep}{1em}
    \centering
    \caption{Example Ledger Rewrites in Cosmos and Ethereum}
    \begin{adjustbox}{width={\textwidth},totalheight={\textheight},keepaspectratio}
    \begin{tabular}[t!]{| p{0.2\linewidth} | p{0.5\linewidth} | p{0.3\linewidth} |}
         \hline
          \textbf{Incident}& \textbf{Description}& \textbf{Impact} \\
         \hline
         \raggedright Juno Proposal 16 & A targeted governance proposal to seize an entity's funds, valued at ~\$100M~\cite{coindesvoteaway}~\cite{coindeskrevoketokens}~\cite{coindeskprop16watershed} & Funds were seized and locked in a smart contract \\
         \hline
         \raggedright Osmosis clawbacks& Two governance proposals to mass alter balances and `claw back' unclaimed airdropped tokens~\cite{mintscanosmosisprop29}~\cite{mintscanosmosisprop32} & Unclaimed airdropped OSMO and ION tokens were clawed back to the chain's Community Pool. \\
         \hline
         \raggedright Tombstone reversions & On the Chihuahua and Persistence networks, validators were tombstoned (a protocol-enforced penalization that permanently deletes both the node operator and delegators' tokens, and permanently removes that validator from the active set of nodes; see Section 4.4) for byzantine behaviour, and sought to have their penalization overturned via governance (See Appendix F, Fig. \ref{fig:q11})~\cite{mintscanpersistenceprop10}~\cite{mintscanhuahuaprop39} & The validators were re-instated, and ledger state altered to restore pre-slash balances. \\
         \hline
         \raggedright Ethereum post-DAO hack rewrite & After a hack, off-chain governance elected to rewrite the chain's state via hard-fork & The chain forked, becoming two different chains, Ethereum and Ethereum Classic. \\
         \hline
    \end{tabular}
    \end{adjustbox}
    \label{table:rewrites}
    }}
\end{table}
It is worth reiterating that although variables in both network topology and governance topology (see Table \ref{table:network_topology} and \ref{table:governance_topology}) can affect resilience to rewrites, decentralization on either axis is independent of resilience. As we hypothesised in Table \ref{table:leviathan_table_1}, dispersion of decision-making power does not prevent rewrites from happening. Indeed, immutability as a design goal might be contrary to any effort at decision-making dispersion, where stakeholders must make nuanced decisions that the protocol cannot handle ex-ante, as this discussion shows,

\blockquote
``Yeah [we have run code that has resulted in a rewrite]\ldots it's controversial in the blockchain world. We think that it's an immutable ledger, but sometimes, shit happens, sometimes [there is a rewrite] for a good reason, like a hacker hacked some account, we want to take it away\ldots The community has a discussion about it and then different people make up their mind, and then [as a validator] our responsibility, once it's passed, we just upgrade the chain, right? If you disagree, you can always exit. It's permissionless.''\ldots

[Prompt: is immutability a core assumption of the technology?] ``We never take an absolutist approach. Everything has a limit\ldots In principle, blockchain is supposed to be immutable, but in some cases we carve out exceptions. Those exceptions will evolve.'' (Interview participant)

\blockquoteendnoskip

\subsection{Agents in the System}
An enumeration and analysis of agents in the system is necessary to understand the governance topology of these networks. In Table \ref{table:agents}, we define five personas, Foundation, Core Team, Contributors, Validators and Stakers. We omit users, as in (D)PoS, stakers are a subset of users who directly interact with the incentive mechanisms of economic finality and chain governance via staking.
{\renewcommand{\arraystretch}{1.5}
\quad{\setlength{\tabcolsep}{1em}
\begin{table}[t!]
    \centering
    \caption{Agents in a typical (Delegated) Proof-of-Stake blockchain}
    \label{table:agents}
    \begin{adjustbox}{width={\textwidth},totalheight={\textheight},keepaspectratio}
    \begin{tabular}{|p{0.2\textwidth}|p{0.2\textwidth}|p{1.4\textwidth}|}\hline
         \textbf{Agent type}&  \raggedright \textbf{Short description}& \textbf{Persona definition}\\\hline
         Foundation&  \raggedright Catch-all term for a central control entity.& For many blockchain projects, there is an entity set up to execute core business, development, and hold tokens, manage payroll et cetera. Since a Foundation in the Caymans is commonly used, the colloquialism Foundation has come to be used for any formal entity nominally in charge of stewarding a blockchain project; ``these foundations are\ldots created in Switzerland or other crypto-friendly countries,''~\cite{mcshanecryptofoundation} and ``are like governments with respect to a national or regional economy, i.e. raising a public treasury and allocating resources to blockchain specific capital works, education, R\&D, etc., to benefit the community and develop the ecosystem.''~\cite{pottsetal}

A more blunt definition can be found in our transcripts, ``[y]ou're trying to distinguish between two organisations. The development team and the Foundation. A blockchain Foundation is what I believe is a quasi-comical legal attempt to separate diverse executives and the lion's share of economic value in the form of tokens and fiat currency from the actual programmatic operations of the blockchain itself.'' (Interview participant)\\\hline
         Core Team&  \raggedright The wider control entity and operations team, which in many cases will include the developers of the software.&Where there is no formal Foundation, Core Team refers to the same stewardship group. This can depend on the stage of the project, “at the early stage of a project, that project [may not] even have a Foundation yet. Foundations [are] typically established at the beginning of the mainnet launch. That's when the project is getting serious, then they want a separate two entities. The Foundation is [say] a Swiss entity, and the Core Team can be a US team.” (Interview participant) Sometimes, in the situation where a Foundation has a more complex structure there might be a distinction between a Foundation that holds tokens at genesis and a company or core team that work on the roadmap for the blockchain. The codebase will normally be open source software, since node operators are expected to check the integrity of software before it is run (see Figure \ref{fig:q7}). In the Cosmos Ecosystem, every major chain is open source, as can be seen from the repository references in the Chain Registry directory.~\cite{cosmoschainregistry} \\\hline
         Contributors&  \raggedright Other developers of the software and project contributors; this will also include open-source contributors who have no formal affiliation with the maintainers.& Often the developers are part of any Core Team, but due to the open-source nature of most blockchain projects, contributors are heterogeneous. They range from members of validator teams contributing bugfixes or documentation, to external or contract developers, or even anonymous contributions.

As described by Spithoven~\cite{spithoven} in their analysis of Bitcoin contributions, development is ``centrally coordinate[d]'' and there is no guarantee that a fork will not be malicious. As early as 2016, the tension between the protocol and its implementers was described, ``[There are two] coordination mechanisms: governance by the infrastructure (achieved via the Bitcoin protocol) and governance of the infrastructure (managed by the community of developers and other stakeholders). [There are] invisible politics inherent in these two mechanisms, which together display a highly technocratic power structure. On the one hand, as an attempt to be self-governing and self-sustaining, the Bitcoin network exhibits a strong market-driven approach to social trust and coordination, which has been embedded directly into the technical protocol. On the other hand, despite being an open source project, the development and maintenance of the Bitcoin code ultimately relies on a small core of highly skilled developers who play a key role in the design of the platform.''~\cite{defilippi2016}
\\\hline
         Validators&  \raggedright Entities and organisations that run validator nodes.&A validator is an ``entity in a distributed ledger technology (DLT) system that participates in validation'' (3.99).~\cite{isoblockchain} In most cases, validators check invariants on a block, arrive at consensus on its inclusion, and commit it. Depending on the blockchain under discussion, the barrier to entry, in terms of technical knowledge, required hardware and investment in minimum bonded stake (where relevant) can be widely variable. 

Validators typically run on more than one network, as shown in Figure \ref{fig:q22}, and do not allocate effort evenly, as shown in Figures \ref{fig:q15} and \ref{fig:q18}, with one operator describing the situation bluntly, ``you divide chains into two different categories. One is like a low-stakes shitchain\ldots Then you have these more high-stakes chains, like Solana.'' (Interview participant) Moreover, validators find many chains are not sustainable (Figure \ref{fig:q19}) and have to balance future chain launches against current network commitments (Figure \ref{fig:q21}). \\\hline
         Stakers&  \raggedright Users that delegate stake to validators in order to secure networks.&Staking necessitates locking or bonding tokens to a validator, lending economic security and allowing economic finality to take place. This exposes stakers to the risk of token loss via a hard or soft slash. However, protocol-enforced penalization in the event of equivocation has often been rolled back as a result of social pressure applied via chain governance, for example on the Chihuahua~\cite{mintscanhuahuaprop39} and Persistence~\cite{mintscanpersistenceprop10} Cosmos networks. Though it is controversial, validators and stakers often seek to undo such automatic penalization via governance (Figure \ref{fig:q11}). \\ \hline
    \end{tabular}
    \end{adjustbox}
\end{table}
}}%

In many Proof-of-Stake (PoS) networks, a variant known as Delegated Proof-of-Stake (DPoS) is used.~\cite{Bashir2022} This is the default in the Cosmos Ecosystem. Stakers bond their tokens to a validator in order to earn rewards from inflation or transaction fees. Validators gain voting power, in direct proportion to the percentage of bonded tokens that are bonded to their validator. This is crucial for economic finality, which relies on a percentage of voting power agreeing on the validity of a block, typically 66.7\%. This finality mechanism is faster than the probabilistic finality used by Proof-of-Work (PoW) networks like Bitcoin. 

In the earliest Proof-of-Stake networks, there was no concept of locking or bonding tokens. Ownership in a wallet was the only requirement, which even more explicitly relied on economic alignment for security of finality. A key concern, if somewhat theoretically, was the so-called ``nothing at stake'' problem: that the dominant strategy of validators would be to add blocks to all proposed versions of history. This dilemma is partly resolved by bonding tokens, and then partly by slashing mechanics, which delete bonded tokens on misbehaving nodes.~\cite{buterinslasher} In Cosmos, the penalization mechanism for equivocation is the so-called hard slash, or tombstone.~\cite{sdkdocstombstone}

\subsection{Collusion and Conflict}
That the alignment of these agents might fail has been critiqued many times. Discussing early Proof-of-Stake systems, Poelstra~\cite{Poelstra2015DistributedCF} argues,

\blockquote
``Essentially, the idea behind using Proof-of-Stake as a consensus mechanism is to move the opportunity costs from outside the system to inside the system\ldots rather than depending on the economic inviability of taking control of a history, stakeholders are incentivized to agree on each extension because (a) they are randomly chosen and therefore unlikely to be in collusion, and (b) even if they can collude, they do not want to undermine the system (e.g. by signing many conflicting histories) because they want to recover their stored value when their stake comes unlocked.''

\blockquoteend

It is property \textit{(b)} that secures the ledger in reality, a hypothesis we derive from our fieldwork and the credit models advanced by Kahn et al.~\cite{kahnetal} (see Future Work). Moreover, a more formal proof is available from Budish et al.,~\cite{budish2024economiclimitspermissionlessconsensus} who describe attacks on this economic security mechanism as being ``expensive due to collapse'' (EDTC). That is, only the risk of scorched-earth economic harm to all token-holders deters malicious behaviour. This was suggested in our fieldwork as the main reason collusion between validators against networks does not occur:

\blockquote
``[The reason is] money, I think, in most of the chains we're involved in. There's very few cases where it would make financial sense for validators to do that, and that's usually because of Foundation delegations and so on. You would have to have some monetary reason to [collude], and I can't see the outcome being better than playing the game fairly, to some extent. This is the fundamentals of Proof-of-Stake, really. If you did something to take over a chain, you're going to tank the value of it, so therefore what is the reason for doing that? That's not to say that there aren't reasons for doing it, but I think we've not seen it because there's not been any reason for validators to do it.'' (Appendix E, `Decentralization Theatre', Response 2)

\blockquoteend

This mirrors the hypothesis of Satoshi Nakamoto in the Bitcoin white paper, describing network takeover, ``[i]f a greedy attacker is able to assemble more CPU power than all the honest nodes, he would have to choose between using it to defraud people by stealing back his payments, or using it to generate new coins. He ought to find it more profitable to play by the rules, such rules that favour him with more new coins than everyone else combined, than to undermine the system and the validity of his own wealth.''~\cite{bitcoin_nakamoto}

All of the actors listed in Table \ref{table:agents} might be assumed to have similar goals, but in practice, they are fluid. The most common assumption is likely to be that all will seek to maintain the economic integrity of a given blockchain. However, stakers do not reliably act in their own economic interest. Economic finality in the case of Proof-of-Stake assigns an economic cost to equivocation. In the case of adversarial ledger rewrites driven by stakers, such as that seen on the Juno network with its Proposal 16,~\cite{mintscanjunoprop16} a cost can be seen. However, these price shocks are hard to disambiguate from background noise, and cannot be relied upon as a means to guarantee that misbehaviour will not occur. Moreover, taking the `expensive due to collapse' (EDTC)~\cite{budish2024economiclimitspermissionlessconsensus} argument to its conclusion, stakers might simply decide that the cost of rewrite, however expensive, is worth it.


\subsection{Consortium vs Permissionless Ledgers}

Across all networks, the need for validators and node operators to take profit creates a real-world constraint on what it means to be `permissionless.' It introduces incentives to collusion and creation of barriers to participation, such as Foundations incentivising node operators, threatening agent independence (see Appendix E, `Decentralization Theatre', Response 5). Actor independence is key to the effectiveness of DLT systems, especially applied to validators, ``[t]he more decentralized the participants in the network---in the sense of being free from the power and control of one another---the more censorship resistant is the ledger.''~\cite{Lemieux_2022} All but one interviewee highlighted the importance of Foundation delegations to their business, with one observing, ``[t]hat's one of the challenges you have with these Foundation delegations. Can you be independent?'' In answer to the question, “How often do foundation grants, delegations, or other direct incentives from a blockchain team determine whether it is profitable or not for your business to operate a validator on that blockchain?” 38\% responded ‘always’. If we include ‘usually’, this rises to 76\% (Figure \ref{fig:q17}).

This potential coercion implies that permissioned and permissionless networks are not that dissimilar in governance topology. In a permissioned ledger, an authority always has the power of veto. In a permissionless ledger, in practice, an authority often has the power to coerce key agents. One might go further and say that just as decentralization must be evaluated in a system at runtime, the same is true of the degree to which it is truly `permissionless.' We address this in Table \ref{table:leviathan_table_1}, where in cases of low or medium dispersion of decision-making power, the way in which a change to ledger state could be forced differs only in that on a permissionless ledger, stakers are an additional potential attacking agent.


Several definitions exist for consortium blockchains. ``A permissioned block-chain requires participants to be authorized first and then participate in network operation with revealed identity. The network governance and consensus body can be either the subsidiaries of a single private entity or a consortium of entities.''~\cite{BAMAKAN2020113385} Other definitions of consortium ledgers do not assume visibility of transactions is limited to only the consortium of node operators and permitted users. ``Consortium blockchains are also known as federated blockchains in which the information is presentable for all people, but its change and acceptance is only possible for determined groups.''~\cite{Xiao_2020} In the second definition, being a consortium blockchain and having publicly visible transactions is not mutually exclusive. This suggests Proof-of-Stake, or Delegated Proof-of-Stake, with validators voting user shares, only differ from consortium blockchains in the ability of a staker to obtain and bond their tokens without consulting a central authority.

Compare these definitions to our fieldwork, which found that nodes were selected by Foundations or core teams for inclusion in a genesis block, and that many node operators subsequently found that Foundation delegations were key to their profitability on a given network. One stated that Foundation delegations were “what makes a chain profitable or not. If you don't get a Foundation delegation, it is typically not profitable to validate on that chain.” Another said, “it's probably the single most important thing in terms of profitability.”

\section{Conclusion}

In this paper, we have advanced several definitions of decentralization, providing one that is nearly adequate to describe the commonly-understood and un-rigorous definitions used by those that work in the space: 

``Decentralization, expressed in terms of control of ledger state, is a holistic expression of the governance layer, that is, on-chain voting power, off-chain coercion, and social influence.''

In unpacking the agents in the system, their interactions and the assumptions, about both the protocol and the incentives that are intended to control them, we can see a chaotic equilibrium at play that can be tuned by certain variables.

In light of this analysis, the previous definition is not enough on its own to analyse and quantify whether or not something is decentralized. For that, we must turn to Vergne's Razor, and ask the question, ``decentralization of what?'' This analysis of the common themes, and variables used to quantify `decentralization' in our fieldwork yields a definition of decentralization that operates on two axes, technical and social:

\begin{enumerate}[noitemsep]
    \item Network topology, \textit{the physical structure of the network}.
    \item Governance topology, \textit{the structure of decision making power in the network}.
\end{enumerate}

The key is that unlike other definitions like Gochhayat et al.,~\cite{gochhayat} we do not distinguish the storage layer. We care about ledger state only insofar as it is an output of these two topologies as the system operates. Additionally, unlike our respondents, we make no claims about whether changing decentralization on these axes affects systemic resilience, even if variables within them can do so.

Of the two, it seems likely that governance topology is the greater lever on perceived decentralization. We hypothesise that it acts as a hard counter to any technical attempt at making any public, permissionless system immutable via its network topology, but definitively proving this is left as future work.

\section{Future Work}

In this paper, we have mentioned the work of Budish,~\cite{budish10.1093/qje/qjae033} and Budish et al.~\cite{budish2024economiclimitspermissionlessconsensus} on the fallibility of economic security in permissionless DLT systems. We have arrived at a similar analysis from our fieldwork, case studies from the Cosmos Ecosystem and an extrapolation of the work of Kahn et al.~\cite{kahnetal} on reversion to autarky in systems of exchange. We intend to build upon this and classify the endogenous tokens discussed in this paper. 


\bibliographystyle{splncs04}
\bibliography{biblio}

\begin{thebibliography}{10}
\providecommand{\url}[1]{\texttt{#1}}
\providecommand{\urlprefix}{URL }
\providecommand{\doi}[1]{https://doi.org/#1}

\bibitem{Axelsen_2022}
Axelsen, H., Jensen, J., Ross, O.: When is a dao decentralized? Complex Systems
  Informatics and Modeling Quarterly (31),  51–75 (Jul 2022).
  \doi{10.7250/csimq.2022-31.04},
  \url{http://dx.doi.org/10.7250/csimq.2022-31.04}

\bibitem{BAMAKAN2020113385}
Bamakan, S.M.H., Motavali, A., {Babaei Bondarti}, A.: {A survey of blockchain
  consensus algorithms performance evaluation criteria}. Expert Systems with
  Applications  \textbf{154},  113385 (2020).
  \doi{https://doi.org/10.1016/j.eswa.2020.113385},
  \url{https://www.sciencedirect.com/science/article/pii/S0957417420302098}

\bibitem{Bashir2022}
Bashir, I.: {{Blockchain Age Protocols}}, pp. 331--376. Apress, Berkeley, CA
  (2022). \doi{10.1007/978-1-4842-8179-6_8},
  \url{https://doi.org/10.1007/978-1-4842-8179-6_8}

\bibitem{budish10.1093/qje/qjae033}
Budish, E.: {Trust at Scale: The Economic Limits of Cryptocurrencies and
  Blockchains}. The Quarterly Journal of Economics  \textbf{140}(1),  1--62 (10
  2024). \doi{10.1093/qje/qjae033}, \url{https://doi.org/10.1093/qje/qjae033}

\bibitem{budish2024economiclimitspermissionlessconsensus}
Budish, E., Lewis-Pye, A., Roughgarden, T.: {The Economic Limits of
  Permissionless Consensus} (2024), \url{https://arxiv.org/abs/2405.09173}

\bibitem{buterinslasher}
Buterin, V.: {{Slasher: A Punitive Proof-of-Stake Algorithm}},
  \url{https://blog.ethereum.org/2014/01/15/slasher-a-punitive-proof-of-stake-algorithm}
  [Accessed: 01.09.2024]

\bibitem{defilippi2016}
De~Filippi, P., Loveluck, B.: {The invisible politics of Bitcoin: governance
  crisis of a decentralised infrastructure}. Internet Policy Review: Journal on
  Internet Regulation  \textbf{5}(3),  1--28 (2016). \doi{10.14763/2016.3.427},
  \url{https://ideas.repec.org/a/zbw/iprjir/214022.html}

\bibitem{mica}
{European Union}: {{REGULATION (EU) 2023/1114 OF THE EUROPEAN PARLIAMENT AND OF
  THE COUNCIL of 31 May 2023 on markets in crypto-assets, and amending
  Regulations (EU) No 1093/2010 and (EU) No 1095/2010 and Directives 2013/36/EU
  and (EU) 2019/1937}},
  \url{https://eur-lex.europa.eu/legal-content/EN/TXT/PDF/?uri=CELEX:32023R1114}
  [Accessed: 14.05.2025]

\bibitem{gochhayat}
Gochhayat, S., Shetty, S., Mukkamala, R., Foytik, P., Kamhoua, G., Njilla, L.:
  {Measuring Decentrality in Blockchain Based Systems}. IEEE Access  \textbf{8}
  (11 2020). \doi{10.1109/ACCESS.2020.3026577}

\bibitem{cosmoschainregistry}
{Interchain Foundation}: {{Cosmos Chain Registry}},
  \url{https://github.com/cosmos/chain-registry} [Accessed: 01.09.2024]

\bibitem{sdkdocstombstone}
{Interchain Foundation}: {{Staking Tombstone}},
  \url{https://docs.cosmos.network/v0.45/modules/slashing/07_tombstone.html}
  [Accessed: 01.09.2024]

\bibitem{whatiscosmos}
{Interchain Foundation}: {{{{What is Cosmos?}}}},
  \url{https://web.archive.org/web/20250125090007/https://v1.cosmos.network/intro}
  [Accessed: 25.01.2025]

\bibitem{isoblockchain}
ISO: {{Blockchain and distributed ledger technologies — Vocabulary}},
  \url{https://www.iso.org/obp/ui/en/#iso:std:iso:22739:ed-2:v1:en} [Accessed:
  01.09.2024]

\bibitem{kahnetal}
Kahn, C., McAndrews, J., Roberds, W.: {Money Is Privacy}. International
  Economic Review - INT ECON REV  \textbf{46},  377--399 (05 2005).
  \doi{10.1111/j.1468-2354.2005.00323.x}

\bibitem{coindeskrevoketokens}
Kessler, S.: {{Juno Blockchain Community Officially Votes to Revoke Whale's
  Tokens}},
  \url{https://www.coindesk.com/layer2/2022/04/29/juno-blockchain-community-officially-votes-to-revoke-whales-tokens/}
  [Accessed: 01.09.2024]

\bibitem{Lemieux_2022}
Lemieux, V.L.: Searching for Trust: Blockchain Technology in an Age of
  Disinformation. Cambridge University Press (2022)

\bibitem{linetal9438771}
Lin, Q., Li, C., Zhao, X., Chen, X.: {Measuring Decentralization in Bitcoin and
  Ethereum using Multiple Metrics and Granularities}. In: {2021 IEEE 37th
  International Conference on Data Engineering Workshops (ICDEW)}. pp. 80--87
  (2021). \doi{10.1109/ICDEW53142.2021.00022}

\bibitem{lynham2025definingdltimmutabilityqualitative}
Lynham, A., Goodell, G.: Defining dlt immutability: A qualitative survey of
  node operators (2025), \url{https://arxiv.org/abs/2507.02413}

\bibitem{mcshanecryptofoundation}
McShane, G.: {{What Is a Crypto Foundation?}},
  \url{https://www.coindesk.com/learn/what-is-a-crypto-foundation/} [Accessed:
  01.09.2024]

\bibitem{mintscanhuahuaprop39}
{Mintscan}: {{Chihuahua Proposal 39: Upgrade v3.1.0 - revert tombstone}},
  \url{https://www.mintscan.io/chihuahua/proposals/39} [Accessed: 01.09.2024]

\bibitem{mintscanjunoprop16}
{Mintscan}: {{Juno Proposal 16: Correcting the gamed stakedrop}},
  \url{https://www.mintscan.io/juno/proposals/16} [Accessed: 01.09.2024]

\bibitem{mintscanosmosisprop29}
{Mintscan}: {{Osmosis Proposal 29: Would you be open to a clawback of
  'unclaimed' ION?}}, \url{https://www.mintscan.io/osmosis/proposals/29}
  [Accessed: 01.09.2024]

\bibitem{mintscanosmosisprop32}
{Mintscan}: {{Osmosis Proposal 32: Clawback ALL unclaimed ION \& OSMO to
  Community Pool after airdrop decay period?}},
  \url{https://www.mintscan.io/osmosis/proposals/32} [Accessed: 01.09.2024]

\bibitem{mintscanpersistenceprop10}
{Mintscan}: {{Persistence Proposal 10: Signalling proposal to revert the
  Tombstone and Slashing from Nov 6th block \#8647535 network halt and restart
  due to appHash mismatch event}},
  \url{https://www.mintscan.io/persistence/proposals/10} [Accessed: 01.09.2024]

\bibitem{coindeskprop16watershed}
Morris, D.: {{Juno's Proposal 16 Vote is a Watershed for Blockchain Governance,
  for Better or Worse}},
  \url{https://www.coindesk.com/layer2/2022/03/16/junos-proposal-16-vote-is-a-watershed-for-blockchain-governance-for-better-or-worse/}
  [Accessed: 01.09.2024]

\bibitem{coindesvoteaway}
Morris, D.: {{We Can Vote Away Your Money For Free}},
  \url{https://www.coindesk.com/layer2/2022/03/11/juno-prop-16-we-can-vote-away-your-money-for-free/}
  [Accessed: 01.09.2024]

\bibitem{bitcoin_nakamoto}
Nakamoto, S.: Bitcoin: A peer-to-peer electronic cash system. Cryptography
  Mailing list at https://metzdowd.com  (03 2009)

\bibitem{edi_ovezik2024sokstratifiedapproachblockchain}
Ovezik, C., Karakostas, D., Kiayias, A.: Sok: A stratified approach to
  blockchain decentralization (2024), \url{https://arxiv.org/abs/2211.01291}

\bibitem{Poelstra2015DistributedCF}
Poelstra, A.: {Distributed Consensus from Proof of Stake is Impossible}. In: {A
  Treatise on Altcoins} (2015),
  \url{https://api.semanticscholar.org/CorpusID:53311056}

\bibitem{pottsetal}
Potts, J., Allen, D., Berg, C., Davidson, S., Macdonald, T.: {An economic
  theory of blockchain foundations}. SSRN Electronic Journal  (01 2021).
  \doi{10.2139/ssrn.3842281}

\bibitem{spithoven}
Spithoven, A.: {Theory and Reality of Cryptocurrency Governance}. Journal of
  Economic Issues  \textbf{53},  385--393 (04 2019).
  \doi{10.1080/00213624.2019.1594518}

\bibitem{srinivasan_lee}
Srinivasan, B. \&~Lee, L.: {{Quantifying Decentralization}},
  \url{https://news.earn.com/quantifying-decentralization-e39db233c28e}
  [Accessed: 01.09.2024]

\bibitem{edi}
{University of Edinburgh}: {{Edinburgh Decentralisation Index}},
  \url{https://blockchainlab.inf.ed.ac.uk/edi-dashboard/} [Accessed:
  14.05.2025]

\bibitem{hhi_antitrust}
{US Department of Justice Antitrust Division}: {{Herfindahl-Hirschman Index}},
  \url{https://web.archive.org/web/20250710215352/https://www.justice.gov/atr/herfindahl-hirschman-index}
  [Accessed: 10.07.2025]

\bibitem{vergne}
Vergne, J.P.: {Web3 as Decentralization Theater? A Framework for Envisioning
  Decentralization Strategically}. Research in the Sociology of Organizations
  \textbf{89},  115--127 (07 2024). \doi{10.1108/S0733-558X20240000089010}

\bibitem{Xiao_2020}
Xiao, Y., Zhang, N., Lou, W., Hou, Y.T.: {A Survey of Distributed Consensus
  Protocols for Blockchain Networks}. IEEE Communications Surveys \& Tutorials
  \textbf{22}(2),  1432–1465 (2020). \doi{10.1109/comst.2020.2969706},
  \url{http://dx.doi.org/10.1109/COMST.2020.2969706}

\end{thebibliography}

\section{Appendix}

\appendix

In this Appendix is a link to the Cosmos Node Data already referenced, the full survey results, as well as longer relevant excerpts from our semi-structured interview transcripts.

\section{Node Data}

Node data including both raw node dumps and screenshots from the Cosmos block explorer Mintscan is available at \href{https://github.com/envoylabs/cosmos\_data/}{this GitHub repository}.

\section{Network and Governance Topology Example Variables}

Example variables identified in our fieldwork are enumerated in tables \ref{table:network_topology} and \ref{table:governance_topology}.

{\renewcommand{\arraystretch}{1.5}
\quad{\setlength{\tabcolsep}{1em}
\begin{table}[H]
    \centering
    \caption{Non-exhaustive, example network topology variables}\label{table:network_topology}
    \begin{adjustbox}{width={\textwidth},totalheight={\textheight},keepaspectratio}
    \begin{tabular}{|p{0.5\linewidth}|p{0.5\linewidth}|}
         \hline
         \textbf{Variable}&\textbf{How to measure this}\\
         \hline
         Number of client implementations& Open-source code repositories\\
         \hline
         Number of active developers& Open-source code repositories\\
         \hline
         Number of commits per active developer& Open-source code repositories\\
         \hline
         Number of validators& Node data or chain explorer\\
         \hline
         Node operator jurisdictions& Data exposed to the network or survey\\
         \hline
         Node data centre distribution& Data exposed to the network or survey\\
         \hline
         Infrastructure service provider& Data exposed to the network or survey\\
         \hline
         Number of archive nodes& Chain explorer or investigation\\
         \hline
    \end{tabular}
    \end{adjustbox}
\end{table}
}}

{\renewcommand{\arraystretch}{1.5}
\quad{\setlength{\tabcolsep}{1em}
\begin{table}[H]
    \centering
    \caption{Non-exhaustive, example governance topology variables}\label{table:governance_topology}
    \begin{adjustbox}{width={\textwidth},totalheight={\textheight},keepaspectratio}
    \begin{tabular}{|p{0.5\linewidth}|p{0.5\linewidth}|}
         \hline
         \textbf{Variable}&\textbf{How to measure this}\\
         \hline
         Permissioned, Consortium or permissionless ledger & Project whitepaper\\
         \hline
         Distribution of token holding& Node data or chain explorer\\
         \hline
         Distribution of staked tokens& Node data or chain explorer\\
         \hline
         On-chain governance vote participation& Node data or chain explorer\\
         \hline
         Vote share of agent category in governance vote& Node data or chain explorer\\
         \hline
         Number of proposers of governance proposals&Node data or chain explorer\\
         \hline
         Number of social media interactions with a governance proposal by stakeholders& Data collection\\
         \hline
 Number of core team members in trusted contact with validator set&Survey\\\hline
    \end{tabular}
    \end{adjustbox}
\end{table}
}}

\section{Semi-structured Interview discussion areas}

\begin{enumerate}[noitemsep]

\item How confident do you feel about the security and robustness of communications related to upgrades, security issues etc?
\item How often do you or your node operations team check the code contained in an upgrade before applying it?
\item Would you or your node operations team knowingly run code that caused a ledger rewrite that affected a single account balance?
\item Have you or your node operations team knowingly run code that caused a ledger rewrite that affected a single account balance?
\item What do you understand by the term ``decentralization''?
\item To what extent do you agree with the following statement: ``There is adequate decentralization on most of the networks we work on.''
\item What do you understand by the term ``decentralization theatre''?
\item Does your organisation validate or mine on any networks that are nominally competitors? What issues can arise?
\item Has your organisation ever experienced a conflict of interest between two networks your organisation validates or mines on?
\item Has your organisation run a node in the genesis set for a permissionless ledger?
\item How did your organisation get selected for that genesis set?
\item If your organisation has run a node in more than one, is there a general pattern for how your organisation has been selected?
\item As professional operators, how concerned are you about privacy on the public networks you work on?
\item What is your opinion on the level of maturity in blockchain governance?

\end{enumerate}

\section{Semi-structured Interviews - Decentralization}

These are responses from our semi-structured interviews to the question, ``What do you understand by the term `decentralization'?'' Note that the response numbers are arbitrarily assigned.

\subsection{Response 1}

The idealistic definition is that there is no central [control], there is no one entity that can control the state of information being held by the blockchain, so that there's not a risk of an authoritarian coming in and rewriting history of some sort. So through decentralization it requires multiple parties to come together and agree at some threshold to provide consensus on a direction that a chain should go. In a nutshell, it requires multiple entities to work together to write the chain. 

[Prompt: how would you distinguish working together from collusion, if it's all one type of entity, for example validators?] There's typically nothing to stop collusion in that manner. The expectation is that for somebody to practically collude at that level on a blockchain that has a large number of participants, that the difficulty of the collusion continues to grow as the blockchain grows. However, if the token controlling interest is held by a select few, sort of like a dev team, and they have a Foundation that is delegating votes, then in effect they have the controlling interest. I can't say that this happens or not, but there certainly could be scenarios in which that's actually what happens, and there could be collusion, yes. 

[Prompt: collusion between the valset against the interests of the Foundation, or between the validators with delegations and the Foundation?] I would say that whoever has the ability to create the initial set of tokens and decide on their distribution has the ability to enact some form of collusion, unless it can be proven that the ownership is in fact distributed and decentralized. Otherwise, it's highly plausible that the Foundation could effectively serve as the authoritarian in that instance. Now, collusion amongst individual validators, there are of course situations where one validator is not actually run by an entity, they are being run on behalf of another entity, so it looks like they are a unique validator, there could actually be one group running ten validators with different names, and there's not really a good way to suss that out and determine who is actually behind it. 

[Prompt: a Sybil attack?] Yes, exactly. 

[Prompt: what is a Foundation in this case?] A Foundation is some sort of steering committee or group whose charter is to foster growth for the blockchain, by creating developer programmes, or marketing, or outreach, or conferences. So for the creation of this Foundation, they're often given a large portion of ownership of the tokens and then they can use them as they see fit to try to achieve their charter.

\subsection{Response 2}

It's hard to define it, but I think I see it from two spectrums. I hope that will be helpful for this question. One is sort of, it's a spectrum in terms of how decentralized something is. If it's a single party running a sequencer, that's not decentralized. On the other hand, if you have a hundred, two hundred, or a thousand, I think everybody can cut how decentralized is sufficient. I might say a hundred validators is sufficient. Somebody will say, ``no, you have to have solo stakers to do it.'' I guess I'm flexible. The other dimension you can look at is a time dimension. A chain can start as rather centralized, and over time you can progressively decentralize. So I don't have a firm definition, I just see decentralization [through] these two lenses, and then different chains will fall into different categories. I'm very tolerant about anything, if a chain is not decentralized on time and space dimensions, that's fine. We will work with it, and hopefully, over time, it will decentralize. So that's my viewpoint. 

[Prompt: what variables could move something from less decentralized to more decentralized, over time?] If it's a spectrum, it's very hard to say what is decentralized and what is not. For me it's very hard to tell. I can pick apart the decentralization discussion and I feel it's very academic in a way. You want to have some kind of measurement, points system, whatever, to maybe rank order. For me it's, just let it go. We are so early in the industry that it's fine, people can experiment. There might be some chains, which I don't see a lot, I would say that most blockchains today are quite decentralized, even though some people will say, you know, Foundations have twenty percent of stake. Well, that's fine. I'm very laid-back. If it's not the right way to do blockchain, those chains will die. So in the end, I don't really worry about centralized chains versus decentralized chains.

\subsection{Response 3}

In the context of blockchains, it's actually multi-layered, because [there's] geographic decentralization of the validators running a chain, and then decentralization of tokens and voting power. A simple [definition] would be that there's no single party controlling the chain. There's multiple different related parts that are involved in the whole thing, I don't know. If any of them are compromised, then the whole system doesn't work. For example, we've had some data centre outages where some chains have stopped running, because all of the validators, or rather, too many of the validators are there. Hetzner used to be a big vendor where a lot of validators were running, but I can't remember if they were a part of that outage. Cost plays a role [in data centre centralization], but also reliability and how well the vendor is working. For example, Hetzner is really hard to compete with, because all of the infrastructure is run so well, and the price is quite low. So from an economic and performance point of view, everybody should be running there. Many chains [now encourage] validators to run elsewhere, so there is a social pressure that is driving the use of other vendors. If looking purely from an economic perspective, it doesn't make sense to run with a much more expensive service provider.

\subsection{Response 4}

I have a negative view on the term decentralization. In my mind, I have my own fantasy view of decentralization, which is where every validator on a particular blockchain is somehow evenly distributed across all the different countries in the world, so you have one validator in every single country, each validator is run by a different person or company, and that validator is on slightly different hardware. Perhaps it's also a different operating system, and on different fibre-optic carriers. In that sense, you have a highly uncorrelated portfolio of validator nodes which, in theory, statistically reduces the probability of any of those validators being sanctioned or turned off. Or if they do go down, from a natural disaster, you're not really going to be impacted. Now that's my fantasy view of decentralization. There's multiple different dimensions, geographical, operator, hardware, and then internet carrier. In reality, most crypto users are not willing to pay for decentralization. If you run a server in different countries, there are different costs associated with that. So some countries might be cheaper to run a validator in, where others are more expensive. That has a lot to do with the cost of the fibre optics, the cost of the power, the availability of servers, and technological hardware, which you may have to import, and pay import tariffs to get servers into a specific location. All of these things add or reduce the cost. So therefore, validators are not economically incentivised to fully decentralize, because the people that pay them will not actually pay for that. They're just going to pay for the cheapest validator, with the lowest commission. So in practice, decentralization does not occur.

\subsection{Response 5}

The core idea is that it's not run by a select few individuals, or a specific company. I think the level of decentralization varies drastically between chains and ecosystems. I don't know exactly what is the right level of decentralization. I think there's loads of metrics, really. A clear one is the size of the validator set, but none of that matters if the validator set is beholden to the chain team and their delegations and things like that, so they almost follow what the chain team directs them to do. I think [stake concentration] matters, you can end up with five validators at the top of the set who effectively control everything, because they have enough stake to vote against things and control governance, but even when you have the stake well-distributed, you have validators that are beholden to delegations that they've received, either by the community or by the chain teams themselves, the Foundations. In both cases, validators are afraid to make the wrong decision and either get the ire of the community or of the chain team. So I don't think we've got it right, but I don't know what the solution is, either. I think decentralization is very much a scale. 

[Prompt: you used chain team and Foundation interchangeably there, what if anything is the difference?] Usually the teams behind the chain itself, the application or the chain. They in most cases have a very large amount of tokens, and they can use that to either bring people into the validator set, or push them out. In the ecosystem that we're involved in, there are a lot of cases where they [the Foundation and development team] are one and the same, or at least, appear to be. So you have the dev teams working directly with the Foundation, or hired by, the Foundation, or the owners of the chain.

\subsection{Response 6}

I feel like decentralization is what's on the billboard, it's how it is sold. But when you really get to understand these networks, whether it's Ethereum, whether it's Maker, whether it's Aptos, even though nodes can be geographically dispersed, I've found that in most cases the majority of the stake is still owned by a small party that could, in fact, enact any change they want. Like Rune, from Maker [DAO], they have this great community, they have this great calls to action with votes, but if the founder of Maker wanted to make any of those changes, he could just push it forward due to his large stake. When you're introducing people to blockchain, the decentralization is part of the sales tactic, but the less decentralized you realise that it is. Especially when you look at [networks like] Aptos, they had thousands of applications for node runners and we got ours based off previous working experience together. I wouldn't say that's fairly decentralized. It's a myth. I think that geographic [decentralization] is important, as well as software client diversification. The economics are probably the most important, because yes, Aptos for example is a public network, and anyone who acquires a million tokens or so can run a node, but that's seven million dollars to get in the door, so it's not exactly democratized. Economic decentralization I think would be the most important thing to have on an actual decentralized network, I feel like the geographic diversification of servers is kind of table stakes. That's what you need at a minimum. 

[Prompt: going above that minimum, would you have in mind things like voting, or decision making?] I do think [decision making dispersion] is important. If decentralization is achieved, then it is of the utmost importance, but the only network that I've seen that has true decentralization, without somebody pulling the strings behind the scenes, whether that's code updates, code pushes, distribution of stake, would be Bitcoin. If I was picking apart my argument, it'd be that the mining pools are somewhat centralized, and have control over which upgrades they choose to operate, or not, so there still is centralization of power. But I think it's the best example of decentralization that I can think of. When you're talking about it from a governance point of view, say they're making changes to Bitcoin core, with halvenings and such, it's really the mining pools that have the decision whether or not they're going to follow along or not. It's been interesting to watch that over the years.

\subsection{Response 7}

I'm getting jaded, but decentralization is more decentralization of several different aspects to protect against legal [sanctions] or one particular person taking over the network. It's providing a security to the chain. It could be hosting providers, it could be machines all running in the same country, or the same timezone, for example, and it could also be where the citizenship or the actual company is based. So, for example, my company is based in, or registered in country XYZ, it is applicable that country XYZ has these laws. Now, if that country happened to be the US, and eighty percent of the validators are also based in the US, and the US says ``it is now illegal for you to do this," they will shut down. So you have to have that kind of decentralization. Similar to Hetzner deciding that they don't want to run crypto networks - and people still do - but they decide that ``today we are going to shut down all Cosmos nodes running on port blahblahblah,'' so you've got different things in there. For the Proof-of-Stake validation, that's where the decentralization [occurs], it's the technical and legal aspects of it. But that doesn't go far enough [unless] it looks at token decentralization. So you have this network where you say ``we're all decentralized, and we're all independent thinkers,'' and all that kind of stuff. You have to take a step back, and then say, ``there is a potential for influence by a whale,'' or a large whale, to potentially put a sizable chunk on each node and then have those nodes, who are independent, basically be coerced, coerced is too strong a term, be motivated to vote in the same direction that he is. Now it's not just Proof-of-Stake chains, this also happens in any kind of governance, where you get a large player that comes in and wants to change the vote, he will find ways to multiply the effect of his token in various ways so that he will get his desired result. So you might hold two percent of a token, and you can make that act like twenty percent of that token by different things. On Proof-of-Stake networks it could be delegating to people [as their biggest delegator], so they're aware that you own two percent of their revenue. For DAOs it could be by, I can't remember what they're called, convex-type situations on a curve, where you own enough of the convex voting rights to affect curves over the whole thing. In that example you have tokens like a Proof-of-Liquidity type thing, where one token owns another token, and they might own ten percent of that token. Maybe you can influence how that ten percent votes, by only holding a tenth of the value of the source token. It's confusing to talk about, but it's a multiplier.

\section{Semi-structured Interviews - Decentralization Theatre}

The following are responses from our semi-structured interviews to the topic area, ``What do you understand by the term `decentralization theatre'?'' Note that the response numbers are arbitrarily assigned.

\subsection{Response 1}

I feel that decentralization is a sales buzzword that people love to hear in crypto, but no-one will actually do anything to incentivise it. So to me it's a bunch of bullshit. When you say `decentralization' in a room of ten people, where you're the eleventh person, if you could magically zoom into everyone's head, you would see ten different results for what each person thinks the word decentralization means. So it's a very diffuse word, and that's why it is such a powerful sales lever. You can say one word, and it speaks custom to each individual person that hears the word, and it doesn't really matter what it means. But nobody's going to pay for it, I'll tell you that.

\subsection{Response 2} 

I think that there are a number of chains that do this, and all of them do it to some extent, in what I was saying about controlling decisions via stake, and so on. But again, I really think that it's a scale. I think that we would all like to be more decentralized, but that comes with a lot of trade-offs. Because we have a lot of companies and institutions controlling these chains or at least creating these chains and then being a fundamental driving force in them, it's always hard to avoid the fact that they will want to push things in a certain direction, and validators haven't got a lot of ways to push back against that. [There's an asymmetry of power], definitely. 

[Prompt: what's to stop validators colluding as a cartel, to counterweight a foundation?] Money, I think, in most of the chains we're involved in. There's very few cases where it would make financial sense for validators to do that, and that's usually because of Foundation delegations and so on. You would have to have some monetary reason to [collude], and I can't see the outcome being better than playing the game fairly, to some extent. This is the fundamentals of Proof-of-Stake, really. If you did something to take over a chain, you're going to tank the value of it, so therefore what is the reason for doing that? That's not to say that there aren't reasons for doing it, but I think we've not seen it because there's not been any reason for validators to do it.

\subsection{Response 3}

[W]hen people are trying to get the validator set [size] increased, it is really a theatre of decentralization, because more validators in practice doesn't mean voting power being spread out to the new validators. It's still concentrated, and it is actually getting worse and worse over time, by people compounding their rewards. 

[Prompt: because stakers compound their staking rewards with the validator they are already staked to?] Yeah.

\subsection{Response 4}

It means that somebody uses decentralization to achieve some goals that are probably not as good, but because they can wave the decentralization flag, it sounds good. It's like a political move that hides some bad intentions, I think. So when you think about decentralization theatre, you can think about it from two different perspectives. One is from the project perspective, where somebody who controls the chain tries to say ``we are decentralized, it's a blockchain project'' to do some kind of unregulated arbitrage, or whatever. The other one I see quite often is the validators. The validators try and run a good business but couldn't get into a chain [validator set], then they will wave the decentralization flag, saying ``hey, you have to support me, because I'm a small validator, I will help you to decentralize.'' As a result, I can make some money. But to go back to the first topic, I'm fully aware that there's some bad actors, especially in 2021, 2022, De-Fi Summer, everybody wanted to be part of blockchain. They tried to take advantage of it with decentralization theatre. What I meant was, a validator, their ultimate motive is to make a business, but they don't want to say it. It's almost taboo to say that a validator is there to make money, so they will say `I'm here for the mission,' you know, if you delegate to me, it's not about me making money, it's about helping you to decentralize. Marketing to [end users and stakers] and to Foundations, which have a bigger stake.

\subsection{Response 5}

Everybody wants to be decentralized, until it affects their bottom line, their wallet. Then all of a sudden, they'll find different ways. If you have a look at most of the chains, you'll find that there is theory that says every Proof-of-Stake validator has a very similar amount of votes on there. So Thorchain is a good example of this where it's one node, one vote. It sounds great, but what happens then is the node operator can actually run multiple nodes. So it still becomes one person, multiple votes.

[Prompt: Is that a Sybil attack?] It's not necessarily a Sybil, because they are all aware of it, of this thing going on. That's just how you make the economics work on Thorchain, but at the start of it, it was one node, one operator. On other chains, [validators] try to attract a larger stake by going to lower commissions, they will offer zero percent or one percent commissions, trying to get a large component of the voting stake. Once that happens, once you've delegated, people usually forget, so they can then mark the prices up so they can actually get value, so they can get a percent or whatever, or even if they stay at zero, they can start having ten percent or fifteen percent of the actual voting power. Now, this presents challenges for a couple of different reasons. Normally on Proof-of-Stake networks, you need to get 66.7\%, two-thirds of the delegated tokens, to agree with your decision. Then if you get these large players, that own ten percent, they have a lot of voting power to make it so that it's not the community's decision, it's their decision. Typically a node operator could be one person or two or three people that control that node, and they would potentially have a much bigger influence in governance. The reason why is that most [stakers] are looking for revenue, they're not looking at the governance side of things, they go, ``I'd rather pay no fees,'' than paying five percent or ten percent fees, ``so I'm going to go with this guy.'' They don't think about governance or decentralization, because they are only in it for a buck. Which is fair, because most people on the chain are only looking at it that way.

[Prompt: So their incentives and the network's don't align?] It's faulty incentives and I think in economic terms they haven't really dealt with the agent problem. So a manager in a large company is supposed to act in the interests of the large company, but in reality they sometimes act in what is best for the manager. For example if my incentive structure is, as an executive in a large company, my incentive is to do X, and I get a large bonus in doing X, I'm going to do X, because that will get me the most money. Now, it might not be the best thing for the company overall, but when I'm looking at it, I'm looking at my wallet, versus anybody else's. The decentralization theatre, or the decentralization of Proof-of-Stake networks, doesn't take that into account, it assumes that everybody has the best interests of the chain and it's very hard, when you're looking at it, to go ``well, my interests are aligned with what the chain's interests are.'' Everybody says that, but sometimes it's a bit, everybody says that they're aligned in the best interests of the chain, but that might not necessarily be what others think. The problem is where you get large validators especially, coming in, and because they have delegations because of zero percent, or very low fees, or something, they can start exerting undue influence. It might not just be zero percent commissions. What you also find is with airdrops, if you delegate to my validator, you will potentially be given some new token in an airdrop in the future. So ``stick with me in the future and you will get an airdrop'' is the other one you see sometimes. You might see that we have five percent commission, but you will get that refunded, so zero percent [commission] by other means.

[Prompt: So the concern is concentration of stake?] The Terra delegation programme, they looked at these things, and said ``we will only delegate to [validators] in this band of criteria, so if you [have] too much money, or delegations, we aren't super interested in you. You need to be active, and you need to have these other things,'' because they were trying to get a score so that every validator had as equal an amount as possible, so you didn't get these huge validators with fifteen or even ten percent of the stake. They're trying to spread it out. On bigger networks, like the Cosmos Hub, it doesn't happen as often, but when there's smaller networks where there isn't that much volume, or there isn't that much TVL [Total Value Locked], you start getting these players coming in and [getting a high percentage of stake], then one or two can basically run the chain. For a small TVL chain, you might get two or three people that have more than 33.3\% of the vote.

\section{Node Operator Interviews - Survey}

A Node Operator survey was distributed via private channels and threads that are used by core teams and Foundations for communications to their validator sets, as well as via private forums for validator teams. The survey questions are below, followed by the results.

\subsection{Survey questions}

\begin{enumerate}[noitemsep]
\item In a business as usual scenario, via which of the following communication channels would you or your node operations team typically expect to hear of a software upgrade (non-consensus-breaking or soft-fork)?

\begin{enumerate}[noitemsep]
    \item Slack
    \item Telegram
    \item Discord
    \item Other closed instant-message platform
    \item Email
    \item IRC
    \item Forum
    \item News article
    \item PagerDuty subscription/notification
    \item On-chain governance proposal (e.g. via Cosmos SDK x/gov)
    \item Github or other code repository (e.g. via release page)
    \item Other (please specify)
\end{enumerate}

\item In a business as usual scenario, via which of the following communication channels would you or your node operations team typically expect to hear of a software upgrade (consensus-breaking or hard-fork)?

\begin{enumerate}[noitemsep]
    \item Slack
    \item Telegram
    \item Discord
    \item Other closed instant-message platform
    \item Email
    \item IRC
    \item Forum
    \item News article
    \item PagerDuty subscription/notification
    \item On-chain governance proposal (e.g. via Cosmos SDK x/gov)
    \item Github or other code repository (e.g. via release page)
    \item Other (please specify)
\end{enumerate}

\item In yours or your node operations team's experience, which of the following is the most common communication channel for operational updates and instructions (for example, upgrades). Please pick only one:

\begin{enumerate}[noitemsep]
    \item Slack
    \item Telegram
    \item Discord
    \item Other closed instant-message platform
    \item Email
    \item IRC
    \item Forum
    \item News article
    \item PagerDuty subscription/notification
    \item On-chain governance proposal (e.g. via Cosmos SDK x/gov)
    \item Github or other code repository (e.g. via release page)
\end{enumerate}

\item How many employees does your organization have?

\begin{enumerate}[noitemsep]
    \item 1
    \item 2-9
    \item 10-49
    \item 50-249
    \item 250+
\end{enumerate}

\item When communicating with the maintainers responsible for upgrading a chain, how many people do you or your node operations team typically have contact with from the counterparty (foundation, core team, etc)?

\begin{enumerate}[noitemsep]
    \item 1
    \item 1-2
    \item 2-5
    \item 5-10
    \item 10+
\end{enumerate}

\item How do you or your node operations team verify the identity of the communicating individual?

\begin{enumerate}[noitemsep]
    \item I/we do not
    \item Signed messages
    \item Exchanging proofs (e.g. public keys)
    \item Checking via second channel, e.g. via email if notified by IM
    \item Verifying with peers
    \item Checking public channels of the counterparty for related news
    \item Other (please specify)
\end{enumerate}

\item How often do you or your node operations team check the code contained in an upgrade before applying it (please be completely honest about this - responses are only used in aggregate)?

\begin{enumerate}[noitemsep]
    \item Always
    \item Usually
    \item About half the time
    \item Seldom
    \item Never
\end{enumerate}

\item Would you or your node operations team knowingly run code that caused a ledger rewrite that affected a single account balance?

\begin{enumerate}[noitemsep]
    \item Yes
    \item No
    \item Undecided
\end{enumerate}

\item Have you or your node operations team knowingly run code that caused a ledger rewrite that affected a single account balance?

\begin{enumerate}[noitemsep]
    \item Yes
    \item No
\end{enumerate}

\item Would you or your node operations team knowingly run code that caused a ledger rewrite that affected multiple account balances in a targeted way (e.g. a clawback)?

\begin{enumerate}[noitemsep]
    \item Yes
    \item No
    \item Undecided
\end{enumerate}

\item Have you or your node operations team knowingly run code that caused a ledger rewrite that affected multiple account balances in a targeted way (e.g. a clawback)?

\begin{enumerate}[noitemsep]
    \item Yes
    \item No
\end{enumerate}

\item In the event of a hard slash that affected a validator run by your organisation, how likely is it that you would seek to undo the effect on the validator and delegators (e.g. via governance)?

\begin{enumerate}[noitemsep]
    \item Definitely
    \item Very Probably
    \item Probably
    \item Possibly
    \item Probably Not
    \item Definitely Not
\end{enumerate}

\item To what extent do you agree with the following statement: ``There is a commonly agreed-upon definition of decentralization that is understood in permissionless networks.''

\begin{enumerate}[noitemsep]
    \item Strongly Agree
    \item Agree
    \item Undecided
    \item Disagree
    \item Strongly Disagree
\end{enumerate}

\item What do you understand by the term ``decentralization''?

Free text box

\item To what extent do you agree with the following statement: ``There is adequate decentralization on most of the networks we work on.''

\item What do you understand by the term ``decentralization theatre''?

Free text box

\item To what extent do you agree with the following statement: ``Concentration of stake is a risk factor on many Proof-of-Stake networks.''

\begin{enumerate}[noitemsep]
    \item Strongly Agree
    \item Agree
    \item Undecided
    \item Disagree
    \item Strongly Disagree
\end{enumerate}

\item To what extent do you agree with the following statement: ``Some networks are more important to our business than others.''

\begin{enumerate}[noitemsep]
    \item Strongly Agree
    \item Agree
    \item Undecided
    \item Disagree
    \item Strongly Disagree
\end{enumerate}

\item To what extent do you agree with the following statement: ``Some networks we validate are not important to our business.''

\begin{enumerate}[noitemsep]
    \item Strongly Agree
    \item Agree
    \item Undecided
    \item Disagree
    \item Strongly Disagree
\end{enumerate}

\item How often do foundation grants, delegations, or other direct incentives from a blockchain team determine whether it is profitable or not for your business to operate a validator on that blockchain?

\begin{enumerate}[noitemsep]
    \item Always
    \item Usually
    \item About half the time
    \item Seldom
    \item Never
\end{enumerate}

\item To what extent do you agree with the following statement: ``We care equally about all networks we validate.''

\begin{enumerate}[noitemsep]
    \item Strongly Agree
    \item Agree
    \item Undecided
    \item Disagree
    \item Strongly Disagree
\end{enumerate}

\item To what extent do you agree with the following statement: ``Some networks we validate have failed in their mission, yet are still running.''

\begin{enumerate}[noitemsep]
    \item Strongly Agree
    \item Agree
    \item Undecided
    \item Disagree
    \item Strongly Disagree
\end{enumerate}

\item To what extent do you agree with the following statement: ``Some networks we validate should shut down.''

\begin{enumerate}[noitemsep]
    \item Strongly Agree
    \item Agree
    \item Undecided
    \item Disagree
    \item Strongly Disagree
\end{enumerate}

\item To what extent do you agree with the following statement: ``As a business, our focus is on chains we currently validate, not on upcoming networks''

\begin{enumerate}[noitemsep]
    \item Strongly Agree
    \item Agree
    \item Undecided
    \item Disagree
    \item Strongly Disagree
\end{enumerate}

\item How many mainnet networks does your organisation or company validate or mine on?

\begin{enumerate}[noitemsep]
    \item 1
    \item 1-2
    \item 2-5
    \item 5-10
    \item 10+
\end{enumerate}

\item Has your organisation run a node in the genesis set for a permissionless ledger?

\begin{enumerate}[noitemsep]
    \item Yes
    \item No
\end{enumerate}

\item How did your organisation get selected for that genesis set?

Free text box

\item If your organisation has run a node in more than one, is there a general pattern for how your organisation has been selected?

Free text box

\item To what extent do you agree with the following statement: ``Privacy is a concern for us as a node operator on public networks''

\begin{enumerate}[noitemsep]
    \item Strongly Agree
    \item Agree
    \item Undecided
    \item Disagree
    \item Strongly Disagree
\end{enumerate}

\item To what extent do you agree with the following statement: ``Privacy is a concern for users on public networks''

\begin{enumerate}[noitemsep]
    \item Strongly Agree
    \item Agree
    \item Undecided
    \item Disagree
    \item Strongly Disagree
\end{enumerate}

\item To what extent do you agree with the following statement: ``Privacy is a concern for the creators, operators or developers of public networks''

\begin{enumerate}[noitemsep]
    \item Strongly Agree
    \item Agree
    \item Undecided
    \item Disagree
    \item Strongly Disagree
\end{enumerate}

\end{enumerate}

\subsection{Survey Free Text Field - Decentralization}

What do you understand by the term ``decentralization''?

\begin{enumerate}[noitemsep]

\item ``Decentralization is a process where every action is spread out to multiple participants of the process it self [sic], where power lies in each participant to some degree depending on their status, position and knowledge.''

\item ``Enough parties involved that the ability to collude to change state, steal, coerce, manipulate is impractical/impossible.  Enough geographic distribution that network/geographic/acts-of-god situations don't impact liveness or state.''

\item ``I understand it to mean that all entities controlling the given subject are fully independent and can act in whichever way they see fit. This is inherently problematic because technically there is no downside to gaming the system unless the majority decide to punish that action, which in my experience is extremely rare.''

\item ``The possibility of generating a newer block for a single newly introduced actor in the ecosystem with reasonable means and owned stake (hardware + energy or PoS coin ownership) allows entities to freely enter or exit, maintaining the system's open and distributed nature.''

\item ``Having as many nodes geographically and provider diverse as possible on your own bare metal hardware located on private property that you own.''

\item ``Multiple distinct entities have to collude to enforce changes to a systems [sic] rules.''

\item ``I understand that it can really mean anything and is dependent on the personal views of the speaker that uses the word `decentralization'. its just a marketing word that gets thrown around in crypto circles, but rarely do people think about what that means at more than the surface level. As a validator, I think deeply about decentralization and I have come to the conclusion that it is not really necessary. ''

\item ``For us true decentralization is a distributed network not only through distributed VP but also location and hosting providers.''

\item ``A substantial effort is made to avoid the system being dependent on a few centralised entities''

\item ``There is no spoon. The term, IMO is meant to represent the distribution of all aspects of a blockchain to many unrelated and autonomous parties. The categories can include diversity in software, hardware location, hardware providers, hardware itself, operator legal jurisdictions, operator affiliations, source of coordination, native(governance) token holder, token holder geographical distribution, token holder legal jurisdiction, token holder control of large amounts, validator voting power, etc\ldots''

\end{enumerate}

\subsection{Survey Free Text Field - Decentralization Theatre}

What do you understand by the term ``decentralization theatre''?

\begin{enumerate}[noitemsep]

\item ``Having many parties involved, when in fact the state is manipulated by a few.''

\item ``It's an act. In reality, most ``decentralized'' PoS networks are just a bunch of people farming what is essentially free money without care or any interest in the actual platform they are operating on.''

\item ``Might refer to the mimicking of decentralization as per other people's definition''

\item ``Having a decentralized system (decentralized validator set) that is effectively controlled by a single entity (majority voting power and code may be controlled by a single entity)''

\item ``Oh boy. Decentralization as a talking point, but not an actual goal.''

\item ``Cosmos decentralization is a joke. There are a handful of entities controlling most chains, and they're almost all mostly in the same Contabo or Hetzner data centers.''

\item ``I'm not familiar with the term but if i was to guess i was say it is when a chain has a high Nakamoto Coefficient but in reality there is 70\% of the chains [sic] VP running on the same hosting provider as a single point of failure. :)''

\item ``Appearing to be decentralised when in reality the system is either run by or controlled by a select few''

\item ``Making a concerted effort to check the commonly regarded ``decentralisation'' boxes to appear decentralised to the general public, while remaining decidedly centralised through one or many other metrics.''

\item ``It is very similar to the security theather [sic]. In this case there are a lot of implementatioms [sic] to get a feel of decentralized network. Usually this is not the case and these protocols just end up masking centralised entety [sic] or a group.''

\end{enumerate}

\subsection{Quantitative Survey Results}

The figures that follow on the next pages show responses to quantitative questions posed in the survey.

\begin{figure*}[h]
\centering
\begin{tikzpicture}
\begin{axis}[%
width=0.84\textwidth, height=6in,
xbar, area legend, bar width=8pt,
xmin=0, xmax=40,
symbolic y coords={o, gh, gov, pd, news, for, irc, email, im, dis, tg, slk},
  ytick=data,
  yticklabels={{Other (please specify)}, {Github or other code repository}, {On-chain governance proposal}, PagerDuty subscription/notification, News article, Forum, IRC, Email, Other closed instant-message platform, Discord, Telegram, Slack},
y tick label style={align=right,text width=3cm},
xmajorgrids,
axis line style={lightgray},
major tick style={draw=none},
nodes near coords,
point meta=explicit symbolic,
node near coords style={font=\footnotesize,right=1em,pin={[pin distance=1em]180:}},
reverse legend
]
\addplot [
  fill={ibmmagenta},draw=none] 
  coordinates {
    (0,o) [0, (0\%)]
    (1,gh) [1, (3.2\%)]
    (0,gov) [0 (0\%)]
    (4,pd) [0 (0\%)]
    (0,news) [0 (0\%)]
    (1,for) [0 (0\%)]
    (0,irc) [0 (0\%)]
    (1,email) [1, (3.2\%)]
    (0,im) [0 (0\%)]
    (28,dis) [28 (90.3\%)]
    (1,tg) [1, (3.2\%)]
    (0,slk) [0 (0\%)]
  };
\addplot [
  fill={ibmyellow},draw=none] 
  coordinates {
    (0,o) [0 (0\%)]
    (7,gh) [7 (22.6\%)]
    (17,gov) [17 54.8\%)]
    (4,pd) [4, (12.9\%)]
    (0,news) [0 (0\%)]
    (1,for) [1 (3.2\%)]
    (0,irc) [0 (0\%)]
    (14,email) [14 (45.2\%)]
    (0,im) [0 (0\%)]
    (29,dis) [29 (93.5\%)]
    (16,tg) [16 (51.6\%)]
    (13,slk) [13 41.9\%)]
  };
\addplot [
  fill={ibmblue},draw=none] 
  coordinates {
    (1,o) [1 (3.2\%)]
    (8,gh) [8 (25.8\%)]
    (14,gov) [14 (45.2\%)]
    (3,pd) [3 (9.7\%)]
    (0,news) [0 (0\%)]
    (1,for) [1 (3.2\%)]
    (0,irc) [0 (0\%)]
    (10,email) [10 (32.3\%)]
    (0,im) [0 (0\%)]
    (31,dis) [31 (100\%)]
    (16,tg) [16 (51.6\%)]
    (12,slk) [12 (38.7\%)]
  };

\draw [line width=1.5pt] (current axis.south west) -- (current axis.north west);
\addlegendentry{Q3}
\addlegendentry{Q2}
\addlegendentry{Q1}
\end{axis}
\end{tikzpicture}
\caption{\\
Q1. In a business as usual scenario, via which of the following communication channels would you or your node operations team typically expect to hear of a software upgrade (non-consensus-breaking or soft-fork)?\\ 
Q2. In a business as usual scenario, via which of the following communication channels would you or your node operations team typically expect to hear of a software upgrade (consensus-breaking or hard-fork)?\\
Q3. In yours or your node operations team's experience, which of the following is the most common communication channel for operational updates and instructions (for example, upgrades). Please pick only one.}
\label{fig:q1}
\end{figure*}

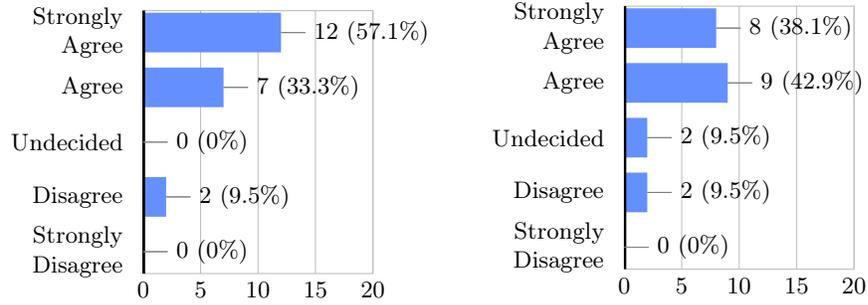
\begin{figure*}[h]
\centering
\begin{tikzpicture}
\begin{axis}[%
width=0.84\textwidth, height=2in,
xbar, bar width=15pt,
xmin=0, xmax=20,
  symbolic y coords={twofiftyplus, fiftyplus, tenfortynine, twonine, one},
  ytick=data,
  yticklabels={{250+}, {50-249}, {10-49}, {2-9}, {1}},
y tick label style={align=right,text width=3cm},
xmajorgrids,
axis line style={lightgray},
major tick style={draw=none},
nodes near coords,
point meta=explicit symbolic,
node near coords style={font=\footnotesize,right=1em,pin={[pin distance=1em]180:}}
]
\addplot [
  fill={ibmblue},draw=none] 
  coordinates {
    (0,twofiftyplus) [0 (0\%)]
    (0,fiftyplus) [0 (0\%)]
    (5,tenfortynine) [5 (16.1\%)]
    (16,twonine) [16 (51.6\%)]
    (10,one) [10 (32.3\%)]
  };
\draw [line width=1.5pt] (current axis.south west) -- (current axis.north west);
\end{axis}
\end{tikzpicture}
\caption{How many employees does your organization have?}
\label{fig:q4}
\end{figure*}
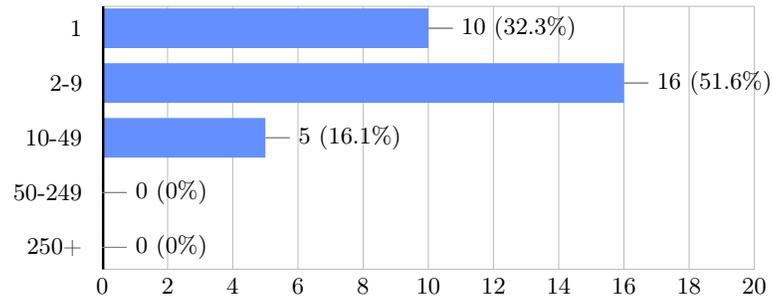

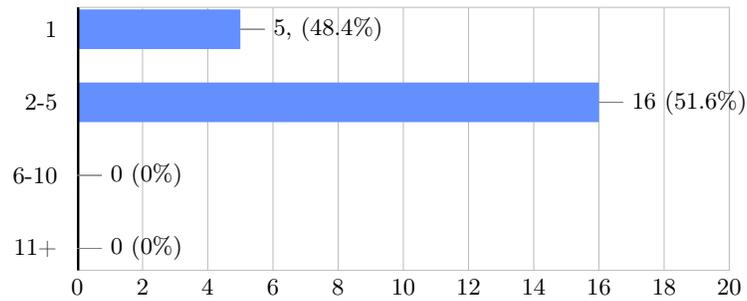
\begin{figure*}[h]
\centering
\begin{tikzpicture}
\begin{axis}[%
width=0.84\textwidth, height=2in,
xbar, bar width=15pt,
xmin=0, xmax=20,
  symbolic y coords={elevenplus, sixten, twofive, one},
  ytick=data,
  yticklabels={{11+}, {6-10}, {2-5}, {1}},
y tick label style={align=right,text width=3cm},
xmajorgrids,
axis line style={lightgray},
major tick style={draw=none},
nodes near coords,
point meta=explicit symbolic,
node near coords style={font=\footnotesize,right=1em,pin={[pin distance=1em]180:}}
]
\addplot [
  fill={ibmblue},draw=none] 
  coordinates {
    (0,elevenplus) [0 (0\%)] 
    (0,sixten) [0 (0\%)]
    (16,twofive) [16 (51.6\%)] 
    (5,one) [5, (48.4\%)] 
  };
\draw [line width=1.5pt] (current axis.south west) -- (current axis.north west);
\end{axis}
\end{tikzpicture}
\caption{When communicating with the maintainers responsible for upgrading a chain, how many people do you or your node operations team typically have contact with from the counterparty (foundation, core team, etc)?}
\label{fig:q5}
\end{figure*}

\begin{figure*}[h]
\centering
\begin{tikzpicture}
\begin{axis}[%
width=0.84\textwidth, height=4in,
xbar, bar width=15pt,
xmin=0, xmax=22,
  symbolic y coords={o,chk,ver,sc,proof,sign,not},
  ytick=data,
  yticklabels={
    {Other (please specify)},
    {Checking public channels of the counterparty for related news},
    {Verifying with peers},
    {Checking via second channel},
    {Exchanging proofs},
    {Signed messages},
    {I/we do not}},
y tick label style={align=right,text width=3cm},
xmajorgrids,
axis line style={lightgray},
major tick style={draw=none},
nodes near coords,
point meta=explicit symbolic,
node near coords style={font=\footnotesize,right=1em,pin={[pin distance=1em]180:}}
]
\addplot [
  fill={ibmblue},draw=none] 
  coordinates {
    (7,o) [7 (22.6\%)]
    (18,chk) [18 (58.1\%)]
    (11,ver) [11 (35.5\%)]
    (9,sc) [9 (29.0\%)]
    (1,proof) [1 (3.2\%)]
    (4,sign) [4 (12.9\%)]
    (3,not) [3 (9.7\%)]
  };
\draw [line width=1.5pt] (current axis.south west) -- (current axis.north west);
\end{axis}
\end{tikzpicture}
\caption{How do you or your node operations team verify the identity of the communicating individual?}
\label{fig:q6}
\end{figure*}
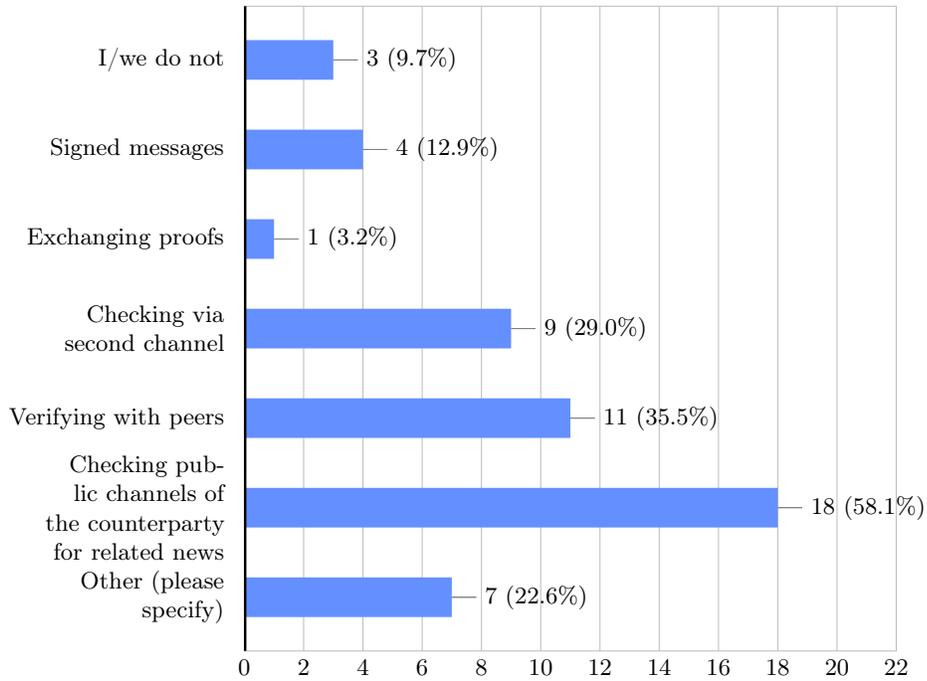

\begin{figure*}[h]
\centering
\begin{tikzpicture}
\begin{axis}[%
width=0.84\textwidth, height=2in,
xbar, bar width=15pt,
xmin=0, xmax=12,
  symbolic y coords={never,seldom,ahtt,usually,always},
  ytick=data,
  yticklabels={
    {Never},
    {Seldom},
    {About half the time},
    {Usually},
    {Always}},
y tick label style={align=right,text width=3cm},
xmajorgrids,
axis line style={lightgray},
major tick style={draw=none},
nodes near coords,
point meta=explicit symbolic,
node near coords style={font=\footnotesize,right=1em,pin={[pin distance=1em]180:}}
]
\addplot [
  fill={ibmblue},draw=none] 
  coordinates {
    (8,never) [8 (25.8\%)]
    (9,seldom) [9 (29.0\%)]
    (5,ahtt) [5 (16.1\%)]
    (4,usually) [4 (12.9\%)]
    (5,always) [5 (16.1\%)]
  };
\draw [line width=1.5pt] (current axis.south west) -- (current axis.north west);
\end{axis}
\end{tikzpicture}
\caption{Q7. How often do [your team] check the code contained in an upgrade before applying it?}
\label{fig:q7}
\end{figure*}
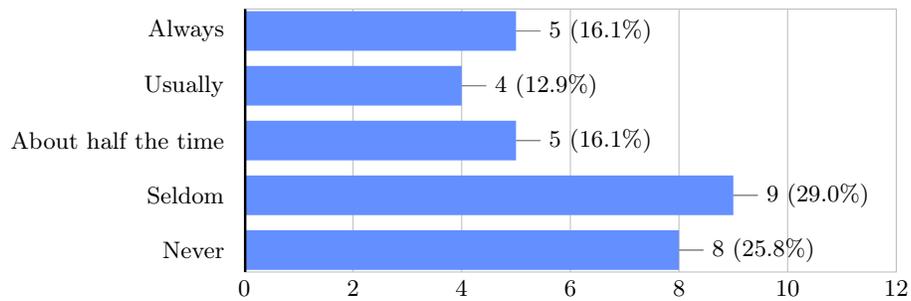

\begin{figure*}[h]
\centering
\begin{tikzpicture}[
    captiontext/.style={below=3mm, text width=5cm}
]

\pie[
    color = {
        ibmmagenta,
    ibmyellow,
    ibmblue}, 
  sum = auto,
    radius = 1.8,
    rotate = 270,
    text=legend
]
    {9/Undecided,
  10/No,
  10/Yes}
\node [captiontext] at (current bounding box.south) {Q9. Would [your team] knowingly run code that caused a ledger rewrite that affected a single account balance?};
\end{tikzpicture}
\begin{tikzpicture}[
    captiontext/.style={below=3mm, text width=5cm}
]

\pie[
    color = {
    ibmyellow,
    ibmblue}, 
  sum = auto,
    radius = 1.8,
    rotate = 270,
    text=legend
    ]
{
    13/No,
  16/Yes}
\node [captiontext] at (current bounding box.south) {Q10. Have [your team] knowingly run code that caused a ledger rewrite that affected a single account balance?};
\end{tikzpicture}
\begin{tikzpicture}[
    captiontext/.style={below=3mm, text width=5cm}
]

\pie[
    color = {
    ibmyellow,
    ibmblue}, 
  sum = auto,
    radius = 1.8,
    rotate = 270,
    text=legend
]{
    14/No,
  15/Yes}
\node [captiontext] at (current bounding box.south) {Q11. Have [your team] knowingly run code that caused a ledger rewrite that affected multiple account balances?};
\end{tikzpicture}
\caption{Validator participation in rewrite events}
\label{fig:q8}
\end{figure*}
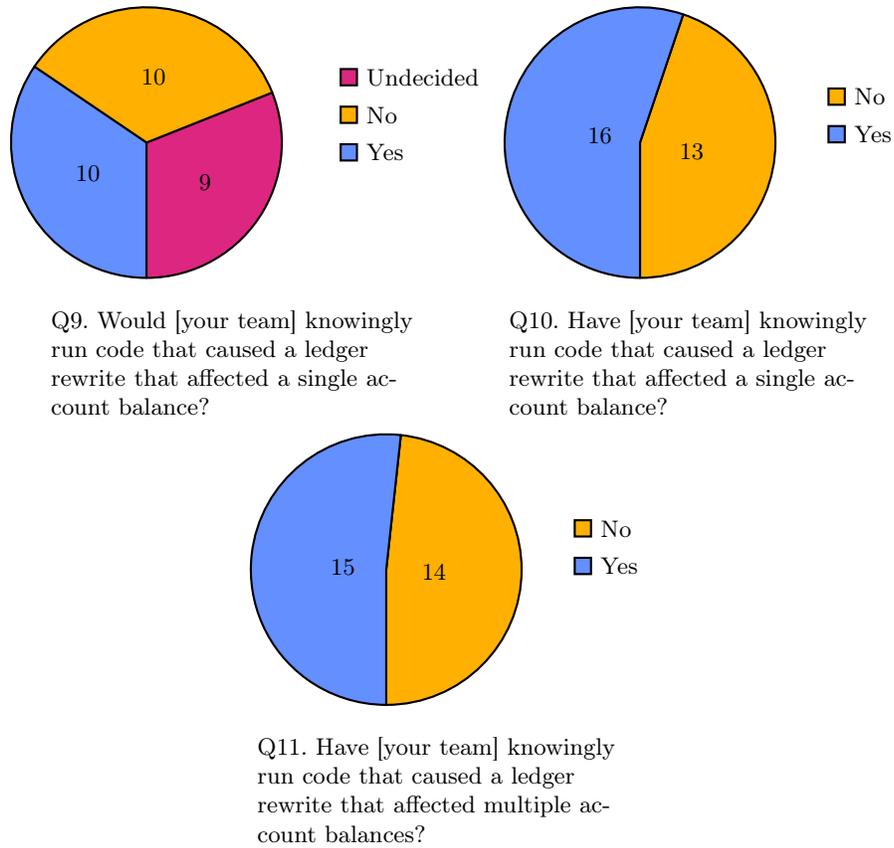

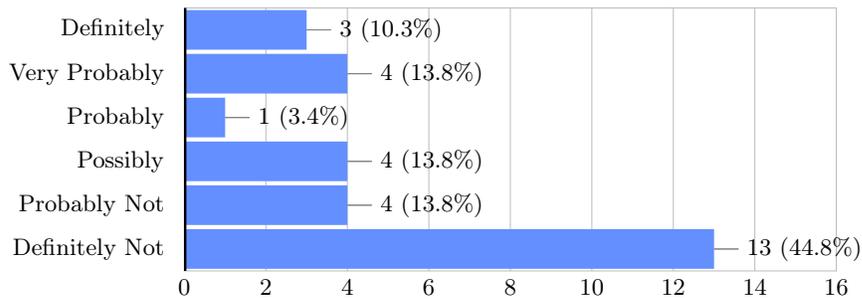
\begin{figure*}[h]
\centering
\begin{tikzpicture}
\begin{axis}[%
width=0.84\textwidth, height=2in,
xbar, bar width=15pt,
xmin=0, xmax=16,
  symbolic y coords={dn,pn,poss,prob,vp,def},
  ytick=data,
  yticklabels={
    {Definitely Not},
    {Probably Not},
    {Possibly},
    {Probably},
    {Very Probably},
    {Definitely}},
y tick label style={align=right,text width=3cm},
xmajorgrids,
axis line style={lightgray},
major tick style={draw=none},
nodes near coords,
point meta=explicit symbolic,
node near coords style={font=\footnotesize,right=1em,pin={[pin distance=1em]180:}}
]
\addplot [
  fill={ibmblue},draw=none] 
  coordinates {
    (13,dn) [13 (44.8\%)]
    (4,pn) [4 (13.8\%)]
    (4,poss) [4 (13.8\%)]
    (1,prob) [1 (3.4\%)]
    (4,vp) [4 (13.8\%)]
    (3,def) [3 (10.3\%)]
  };
\draw [line width=1.5pt] (current axis.south west) -- (current axis.north west);
\end{axis}
\end{tikzpicture}
\caption{Q12. In the event of a hard slash that affected a validator run by your organisation, how likely is it that you would seek to undo the effect on the validator and delegators (e.g. via governance)?}
\label{fig:q11}
\end{figure*}

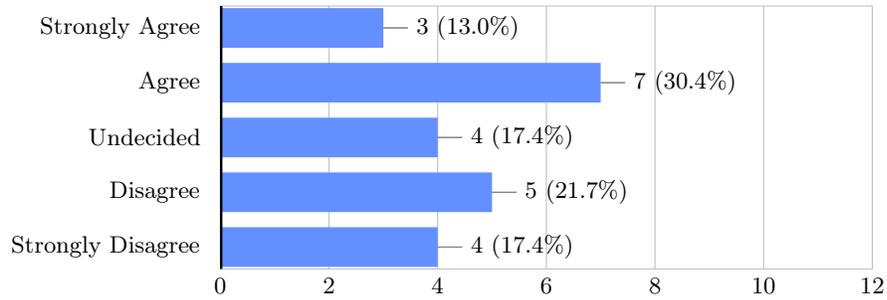
\begin{figure*}[h]
\centering
\begin{tikzpicture}
\begin{axis}[%
width=0.84\textwidth, height=2in,
xbar, bar width=15pt,
xmin=0, xmax=12,
  symbolic y coords={sdisag,disag,un,ag,sag},
  ytick=data,
  yticklabels={
    {Strongly Disagree},
    {Disagree},
    {Undecided},
    {Agree},
    {Strongly Agree}},
y tick label style={align=right,text width=3cm},
xmajorgrids,
axis line style={lightgray},
major tick style={draw=none},
nodes near coords,
point meta=explicit symbolic,
node near coords style={font=\footnotesize,right=1em,pin={[pin distance=1em]180:}}
]
\addplot [
  fill={ibmblue},draw=none] 
  coordinates {
    (4,sdisag) [4 (17.4\%)]
    (5,disag) [5 (21.7\%)]
    (4,un) [4 (17.4\%)]
    (7,ag) [7 (30.4\%)]
    (3,sag) [3 (13.0\%)]
  };
\draw [line width=1.5pt] (current axis.south west) -- (current axis.north west);
\end{axis}
\end{tikzpicture}
\caption{Q13. To what extent do you agree with the following statement: ``There is a commonly agreed-upon definition of decentralization that is understood in permissionless networks.''}
\label{fig:q12}
\end{figure*}

\begin{figure*}[h]
\centering
\begin{tikzpicture}
\begin{axis}[%
width=0.84\textwidth, height=2in,
xbar, bar width=15pt,
xmin=0, xmax=12,
  symbolic y coords={sdisag,disag,un,ag,sag},
  ytick=data,
  yticklabels={
    {Strongly Disagree},
    {Disagree},
    {Undecided},
    {Agree},
    {Strongly Agree}},
y tick label style={align=right,text width=3cm},
xmajorgrids,
axis line style={lightgray},
major tick style={draw=none},
nodes near coords,
point meta=explicit symbolic,
node near coords style={font=\footnotesize,right=1em,pin={[pin distance=1em]180:}}
]
\addplot [
  fill={ibmblue},draw=none] 
  coordinates {
    (9,sdisag) [9 (39.1\%)]
    (6,disag) [6 (26.1\%)]
    (1,un) [1 (4.3\%)]
    (6,ag) [6 (26.1\%)]
    (1,sag) [1 (4.3\%)]
  };
\draw [line width=1.5pt] (current axis.south west) -- (current axis.north west);
\end{axis}
\end{tikzpicture}
\caption{Q15. To what extent do you agree with the following statement: ``There is adequate decentralization on most of the networks we work on.''}
\label{fig:q13}
\end{figure*}
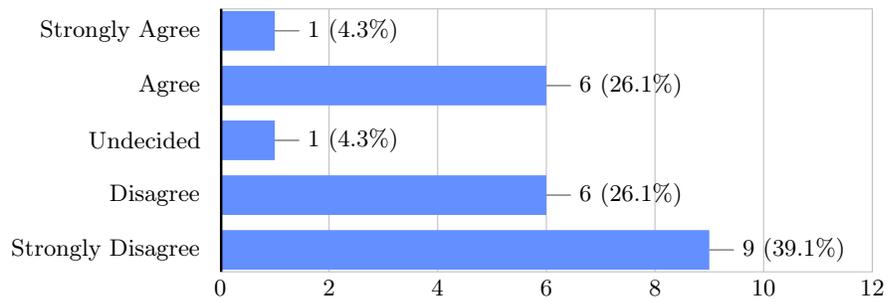

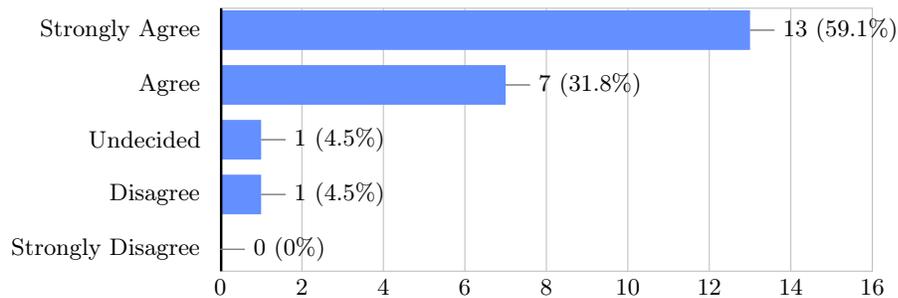
\begin{figure*}[h]
\centering
\begin{tikzpicture}
\begin{axis}[%
width=0.84\textwidth, height=2in,
xbar, bar width=15pt,
xmin=0, xmax=16,
  symbolic y coords={sdisag,disag,un,ag,sag},
  ytick=data,
  yticklabels={
    {Strongly Disagree},
    {Disagree},
    {Undecided},
    {Agree},
    {Strongly Agree}},
y tick label style={align=right,text width=3cm},
xmajorgrids,
axis line style={lightgray},
major tick style={draw=none},
nodes near coords,
point meta=explicit symbolic,
node near coords style={font=\footnotesize,right=1em,pin={[pin distance=1em]180:}}
]
\addplot [
  fill={ibmblue},draw=none] 
  coordinates {
    (0,sdisag) [0 (0\%)]
    (1,disag) [1 (4.5\%)]
    (1,un) [1 (4.5\%)]
    (7,ag) [7 (31.8\%)]
    (13,sag) [13 (59.1\%)]
  };
\draw [line width=1.5pt] (current axis.south west) -- (current axis.north west);
\end{axis}
\end{tikzpicture}
\caption{Q17. To what extent do you agree with the following statement: ``Concentration of stake is a risk factor on many Proof-of-Stake networks.''}
\label{fig:q14}
\end{figure*}

\begin{figure*}[h]
\centering
\begin{tikzpicture}[
    captiontext/.style={below=3mm, text width=5cm}
]
\begin{axis}[%
width=0.38\textwidth, height=2in,
xbar, bar width=15pt,
xmin=0, xmax=20,
  symbolic y coords={sdisag,disag,un,ag,sag},
  ytick=data,
  yticklabels={
    {Strongly Disagree},
    {Disagree},
    {Undecided},
    {Agree},
    {Strongly Agree}},
y tick label style={align=right,text width=2cm},
xmajorgrids,
axis line style={lightgray},
major tick style={draw=none},
nodes near coords,
point meta=explicit symbolic,
node near coords style={font=\footnotesize,right=1em,pin={[pin distance=1em]180:}}
]
\addplot [
  fill={ibmblue},draw=none] 
  coordinates {
    (0,sdisag) [0 (0\%)]
    (2,disag) [2 (9.5\%)]
    (0,un) [0 (0\%)]
    (7,ag) [7 (33.3\%)]
    (12,sag) [12 (57.1\%)]
  };
\draw [line width=1.5pt] (current axis.south west) -- (current axis.north west);
\end{axis}
\node [captiontext] at (current bounding box.south) {Q18. To what extent do you agree with the following statement: ``Some networks are more important to our business than others.''};
\end{tikzpicture}
\begin{tikzpicture}[
    captiontext/.style={below=3mm, text width=5cm}
]
\begin{axis}[%
width=0.38\textwidth, height=2in,
xbar, bar width=15pt,
xmin=0, xmax=20,
  symbolic y coords={sdisag,disag,un,ag,sag},
  ytick=data,
  yticklabels={
    {Strongly Disagree},
    {Disagree},
    {Undecided},
    {Agree},
    {Strongly Agree}},
y tick label style={align=right,text width=2cm},
xmajorgrids,
axis line style={lightgray},
major tick style={draw=none},
nodes near coords,
point meta=explicit symbolic,
node near coords style={font=\footnotesize,right=1em,pin={[pin distance=1em]180:}}
]
\addplot [
  fill={ibmblue},draw=none] 
  coordinates {
    (0,sdisag) [0 (0\%)]
    (2,disag) [2 (9.5\%)]
    (2,un) [2 (9.5\%)]
    (9,ag) [9 (42.9\%)]
    (8,sag) [8 (38.1\%)]
  };
\draw [line width=1.5pt] (current axis.south west) -- (current axis.north west);
\end{axis}
\node [captiontext] at (current bounding box.south) {Q19. To what extent do you agree with the following statement: ``Some networks we validate are not important to our business.''};
\end{tikzpicture}

\caption{Validator business priorities}
\label{fig:q15}
\end{figure*}

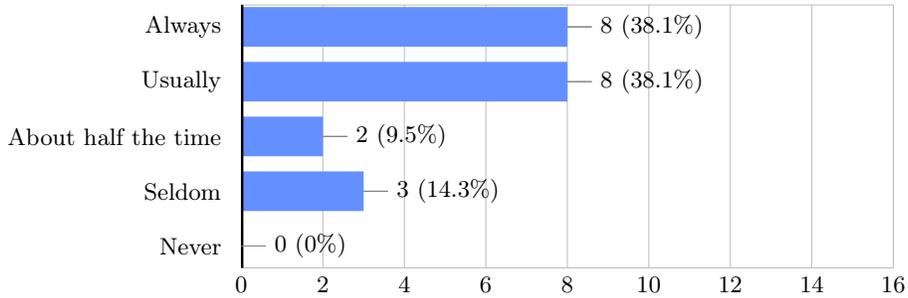
\begin{figure*}[h]
\centering
\begin{tikzpicture}
\begin{axis}[%
width=0.84\textwidth, height=2in,
xbar, bar width=15pt,
xmin=0, xmax=16,
  symbolic y coords={never,seldom,ahtt,usually,always},
  ytick=data,
  yticklabels={
    {Never},
    {Seldom},
    {About half the time},
    {Usually},
    {Always}},
y tick label style={align=right,text width=3cm},
xmajorgrids,
axis line style={lightgray},
major tick style={draw=none},
nodes near coords,
point meta=explicit symbolic,
node near coords style={font=\footnotesize,right=1em,pin={[pin distance=1em]180:}}
]
\addplot [
  fill={ibmblue},draw=none] 
  coordinates {
    (0,never) [0 (0\%)]
    (3,seldom) [3 (14.3\%)]
    (2,ahtt) [2 (9.5\%)]
    (8,usually) [8 (38.1\%)]
    (8,always) [8 (38.1\%)]
  };
\draw [line width=1.5pt] (current axis.south west) -- (current axis.north west);
\end{axis}
\end{tikzpicture}
\caption{Q20. How often do foundation grants, delegations, or other direct incentives from a blockchain team determine whether it is profitable or not for your business to operate a validator on that blockchain?}
\label{fig:q17}
\end{figure*}

\begin{figure*}[h]
\centering
\begin{tikzpicture}
\begin{axis}[%
width=0.84\textwidth, height=2in,
xbar, bar width=15pt,
xmin=0, xmax=16,
  symbolic y coords={sdisag,disag,un,ag,sag},
  ytick=data,
  yticklabels={
    {Strongly disagree},
    {Disagree},
    {Undecided},
    {Agree},
    {Strongly agree}},
y tick label style={align=right,text width=3cm},
xmajorgrids,
axis line style={lightgray},
major tick style={draw=none},
nodes near coords,
point meta=explicit symbolic,
node near coords style={font=\footnotesize,right=1em,pin={[pin distance=1em]180:}}
]
\addplot [
  fill={ibmblue},draw=none] 
  coordinates {
    (4,sdisag) [4 (19.0\%)]
    (10,disag) [10 (47.6\%)]
    (0,un) [0 (0\%)]
    (5,ag) [5 (23.8\%)]
    (2,sag) [2 (9.5\%)]
  };
\draw [line width=1.5pt] (current axis.south west) -- (current axis.north west);
\end{axis}
\end{tikzpicture}
\caption{To what extent do you agree with the following statement: ``We care equally about all networks we validate.''}
\label{fig:q18}
\end{figure*}
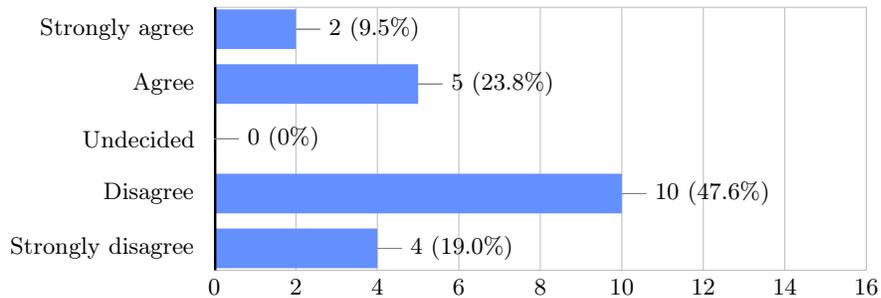

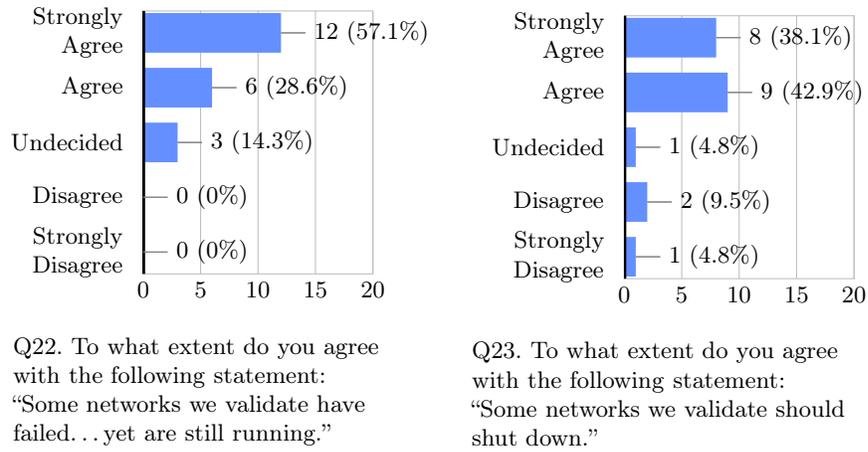
\begin{figure*}[h]
\centering
\begin{tikzpicture}[
    captiontext/.style={below=3mm, text width=5cm}
]
\begin{axis}[%
width=0.38\textwidth, height=2in,
xbar, bar width=15pt,
xmin=0, xmax=20,
  symbolic y coords={sdisag,disag,un,ag,sag},
  ytick=data,
  yticklabels={
    {Strongly Disagree},
    {Disagree},
    {Undecided},
    {Agree},
    {Strongly Agree}},
y tick label style={align=right,text width=2cm},
xmajorgrids,
axis line style={lightgray},
major tick style={draw=none},
nodes near coords,
point meta=explicit symbolic,
node near coords style={font=\footnotesize,right=1em,pin={[pin distance=1em]180:}}
]
\addplot [
  fill={ibmblue},draw=none] 
  coordinates {
    (0,sdisag) [0 (0\%)]
    (0,disag) [0 (0\%)]
    (3,un) [3 (14.3\%)]
    (6,ag) [6 (28.6\%)]
    (12,sag) [12 (57.1\%)]
  };
\draw [line width=1.5pt] (current axis.south west) -- (current axis.north west);
\end{axis}
\node [captiontext] at (current bounding box.south) {Q22. To what extent do you agree with the following statement: ``Some networks we validate have failed\ldots yet are still running.''};
\end{tikzpicture}
\begin{tikzpicture}[
    captiontext/.style={below=3mm, text width=5cm}
]
\begin{axis}[%
width=0.38\textwidth, height=2in,
xbar, bar width=15pt,
xmin=0, xmax=20,
  symbolic y coords={sdisag,disag,un,ag,sag},
  ytick=data,
  yticklabels={
    {Strongly Disagree},
    {Disagree},
    {Undecided},
    {Agree},
    {Strongly Agree}},
y tick label style={align=right,text width=2cm},
xmajorgrids,
axis line style={lightgray},
major tick style={draw=none},
nodes near coords,
point meta=explicit symbolic,
node near coords style={font=\footnotesize,right=1em,pin={[pin distance=1em]180:}}
]
\addplot [
  fill={ibmblue},draw=none] 
  coordinates {
    (1,sdisag) [1 (4.8\%)]
    (2,disag) [2 (9.5\%)]
    (1,un) [1 (4.8\%)]
    (9,ag) [9 (42.9\%)]
    (8,sag) [8 (38.1\%)]
  };
\draw [line width=1.5pt] (current axis.south west) -- (current axis.north west);
\end{axis}
\node [captiontext] at (current bounding box.south) {Q23. To what extent do you agree with the following statement: ``Some networks we validate should shut down.''};
\end{tikzpicture}

\caption{Validator assessment of network success or failure}
\label{fig:q19}
\end{figure*}

\begin{figure*}[h]
\centering
\begin{tikzpicture}
\begin{axis}[%
width=0.84\textwidth, height=2in,
xbar, bar width=15pt,
xmin=0, xmax=16,
  symbolic y coords={sdisag,disag,un,ag,sag},
  ytick=data,
  yticklabels={
    {Strongly disagree},
    {Disagree},
    {Undecided},
    {Agree},
    {Strongly agree}},
y tick label style={align=right,text width=3cm},
xmajorgrids,
axis line style={lightgray},
major tick style={draw=none},
nodes near coords,
point meta=explicit symbolic,
node near coords style={font=\footnotesize,right=1em,pin={[pin distance=1em]180:}}
]
\addplot [
  fill={ibmblue},draw=none] 
  coordinates {
    (3,sdisag) [3 (14.3\%)]
    (7,disag) [7 (33.3\%)]
    (4,un) [4 (19.0\%)]
    (4,ag) [4 (19.0\%)]
    (3,sag) [3 (14.3\%)]
  };
\draw [line width=1.5pt] (current axis.south west) -- (current axis.north west);
\end{axis}
\end{tikzpicture}
\caption{To what extent do you agree with the following statement: ``As a business, our focus is on chains we currently validate, not on upcoming networks''}
\label{fig:q21}
\end{figure*}
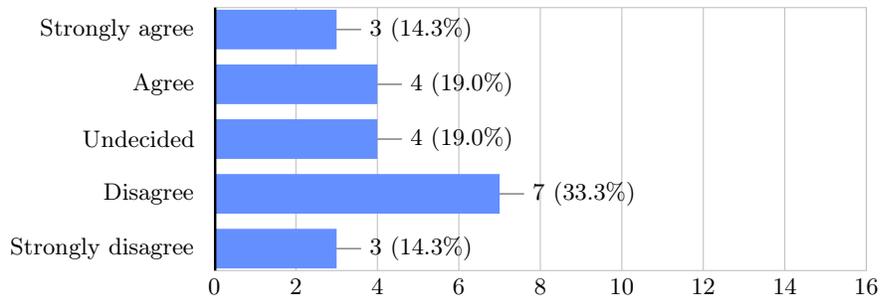

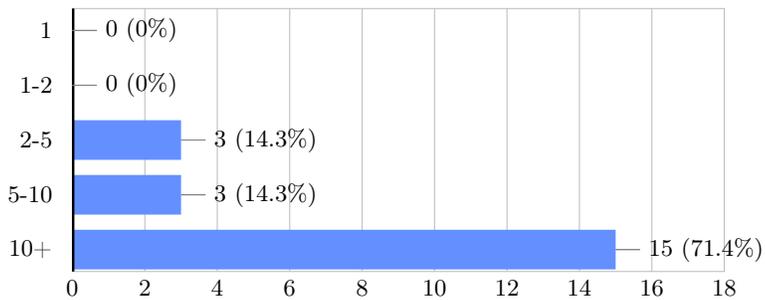
\begin{figure*}[h]
\centering
\begin{tikzpicture}
\begin{axis}[%
width=0.84\textwidth, height=2in,
xbar, bar width=15pt,
xmin=0, xmax=18,
  symbolic y coords={tenplus,fiveten,twofive,onetwo,one},
  ytick=data,
  yticklabels={
    {10+},
    {5-10},
    {2-5},
    {1-2},
    {1}},
y tick label style={align=right,text width=3cm},
xmajorgrids,
axis line style={lightgray},
major tick style={draw=none},
nodes near coords,
point meta=explicit symbolic,
node near coords style={font=\footnotesize,right=1em,pin={[pin distance=1em]180:}}
]
\addplot [
  fill={ibmblue},draw=none] 
  coordinates {
    (15,tenplus) [15 (71.4\%)]
    (3,fiveten) [3 (14.3\%)]
    (3,twofive) [3 (14.3\%)]
    (0,onetwo) [0 (0\%)]
    (0,one) [0 (0\%)]
  };
\draw [line width=1.5pt] (current axis.south west) -- (current axis.north west);
\end{axis}
\end{tikzpicture}
\caption{Q25. How many mainnet networks does your organisation or company validate or mine on?}
\label{fig:q22}
\end{figure*}

\begin{figure*}[ht]
\centering
\begin{tikzpicture}

\pie[
    color = {
    ibmyellow,
    ibmblue}, 
  sum = auto,
    radius = 1.8,
    text=legend
]
    {1/No,
  20/Yes}
\end{tikzpicture}
\caption{Has your organisation run a node in the genesis set for a permissionless ledger?}
\label{fig:q23}
\end{figure*}
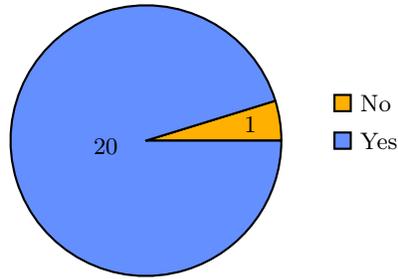

\begin{figure*}[h]
\centering
\begin{tikzpicture}
\begin{axis}[%
width=0.84\textwidth, height=4in,
xbar,area legend, bar width=10pt,
xmin=0, xmax=16,
  symbolic y coords={sdisag,disag,un,ag,sag},
  ytick=data,
  yticklabels={
    {Strongly Disagree},
    {Disagree},
    {Undecided},
    {Agree},
    {Strongly Agree}},
y tick label style={align=right,text width=3cm},
xmajorgrids,
axis line style={lightgray},
major tick style={draw=none},
nodes near coords,
point meta=explicit symbolic,
node near coords style={font=\footnotesize,right=1em,pin={[pin distance=1em]180:}},
reverse legend
]
\addplot [
  fill={ibmmagenta},draw=none] 
  coordinates {
    (1,sdisag) [1 (4.8\%)]
    (5,disag) [5 (23.8\%)]
    (6,un) [6 (28.6\%)]
    (7,ag) [7 (33.3\%)]
    (2,sag) [2 (9.5\%)]
  };
\addplot [
  fill={ibmyellow},draw=none] 
  coordinates {
    (0,sdisag) [0 (0\%)]
    (4,disag) [4 (19.0\%)]
    (4,un) [4 (19.0\%)]
    (11,ag) [11 (52.4\%)]
    (2,sag) [2 (9.5\%)]
  };
  \addplot [
  fill={ibmblue},draw=none] 
  coordinates {
    (0,sdisag) [0 (0\%)]
    (5,disag) [5 (23.8\%)]
    (4,un) [4 (19.0\%)]
    (9,ag) [9 (42.9\%)]
    (3,sag) [3 (14.3\%)]
  };
\draw [line width=1.5pt] (current axis.south west) -- (current axis.north west);
\addlegendentry{Q32}
\addlegendentry{Q31}
\addlegendentry{Q30}
\end{axis}
\end{tikzpicture}
\caption{\\
Q30. To what extent do you agree with the following statement: ``Privacy is a concern for us as a node operator on public networks''\\
Q31. To what extent do you agree with the following statement: ``Privacy is a concern for the creators, operators or developers of public networks''\\
Q32. To what extent do you agree with the following statement: ``Privacy is a concern for the creators, operators or developers of public networks''}
\label{fig:q27}
\end{figure*}



%
%
%

\end{document}